\documentclass[a4paper,reprint,aps,prd,groupedaddress,nofootinbib]{revtex4-2}

\usepackage{amsmath,amssymb,mathtools}
\usepackage{bm}
\usepackage{tensor}
\allowdisplaybreaks

\usepackage{array}[=2016-10-06]
\usepackage{dcolumn}
\usepackage{booktabs}
\usepackage{tabularx}
\usepackage{multirow}

\usepackage{graphicx}
\usepackage{rotating}
\usepackage{adjustbox}

\usepackage{csquotes}
\usepackage[dvipsnames]{xcolor}
\usepackage{hyperref}
\usepackage{orcidlink}
\usepackage[capitalise]{cleveref}

\hypersetup{
  colorlinks=true,
  linkcolor=blue,
  citecolor=MidnightBlue,
  urlcolor=Blue
}

\definecolor{amethyst}{rgb}{0.6,0.4,0.8}
\definecolor{customblue}{rgb}{0.2,0.2,0.8}

\begin{document}

\title{Testing Schouten--Codazzi Gravity: Cosmological Constraints and Deviations from $\Lambda$CDM}

\author{Andoni Fontanet}
\email{andonifon@gmail.com}
\affiliation{Departamento de F\'isica y Matem\'aticas, Universidad Iberoamericana, Prolongaci\'on Paseo de la Reforma 880, 01219 Ciudad de M\'exico, M\'exico}
\author{Andrea Valencia}
\email{andreavalenciar8@gmail.com}
\affiliation{Departamento de F\'isica y Matem\'aticas, Universidad Iberoamericana, Prolongaci\'on Paseo de la Reforma 880, 01219 Ciudad de M\'exico, M\'exico}

\author{Shahnawaz A. Adil}
\email{shazadil14@gmail.com}
\affiliation{Instituto de Ciencias F\'isicas, Universidad Nacional Aut\'onoma de M\'exico, Cuernavaca, Morelos, 62210, M\'exico}
\author{Sebasti\'an N\'ajera}
\email{sebastian.najera@icf.unam.mx}
\affiliation{Instituto de Ciencias F\'isicas, Universidad Nacional Aut\'onoma de M\'exico, Cuernavaca, Morelos, 62210, M\'exico}
\author{Roberto A. Sussman}
\email{sussman@nucleares.unam.mx}
\affiliation{Instituto de Ciencias F\'isicas, Universidad Nacional Aut\'onoma de M\'exico, Cuernavaca, Morelos, 62210, M\'exico}
\author{J. Alberto V\'azquez}
\email{javazquez@icf.unam.mx}
\affiliation{Instituto de Ciencias F\'isicas, Universidad Nacional Aut\'onoma de M\'exico, Cuernavaca, Morelos, 62210, M\'exico}
\author{J. C. Hidalgo}
\email{hidalgo@icf.unam.mx}
\affiliation{Instituto de Ciencias F\'isicas, Universidad Nacional Aut\'onoma de M\'exico, Cuernavaca, Morelos, 62210, M\'exico}

\date{\today}

\begin{abstract}

Schouten--Codazzi gravity (SCG) admits a cosmological background that extends flat $\Lambda$CDM by one integration constant, $\gamma_0$. Its contribution to the Friedmann equation scales as $(1+z)$ and corresponds to an effective component with equation of state $w_{\gamma_0}=-2/3$. We constrain this extension using 15 combinations of Type~Ia supernovae, DESI~DR2 baryon acoustic oscillations, and compressed cosmic microwave background information, supplemented by a big-bang nucleosynthesis prior as specified in the analysis. Supernovae paired with either BAO or CMB favor positive $\Omega_{\gamma_0}\equiv\gamma_0/H_0^2$, with approximate marginal departures from zero of $1.7$--$3.2\sigma$. In contrast, all three supernovae$+$BAO$+$CMB combinations are consistent with $\Omega_{\gamma_0}=0$ within their 68\% credible intervals. Their central log Bayes factors favor $\Lambda$CDM, with $\Delta\ln\mathcal{Z}$ between $-2.54$ and $-1.13$, and the inferred Hubble constants remain close to those of the corresponding $\Lambda$CDM fits. These results provide a background-level test of SCG conditional on the adopted priors and likelihood approximations, particularly the model assumptions entering the CMB compression.
\end{abstract}

\maketitle

\section{Introduction}

The $\Lambda$CDM model provides a simple and successful description of a broad range of cosmological observations. Nevertheless, comparisons among cosmic microwave background (CMB), baryon acoustic oscillation (BAO), large-scale structure (LSS), and local distance-ladder measurements have revealed discrepancies that motivate further tests of its assumptions~\cite{Perivolaropoulos:2021jda,Abdalla:2022yfr,DiValentino:2022fjm,Peebles:2024txt,DiValentino:2025sru}.

As observational precision improves~\cite{DiValentino:2020vhf}, increasingly stringent tests of the standard $\Lambda$CDM picture become possible. In fact, several inconsistencies in the inferred values of central cosmological parameters have already been reported, with some reaching statistically meaningful levels of disagreement with the model's predictions~\cite{Abdalla:2022yfr,DiValentino:2020vvd,DiValentino:2020srs,DiValentino:2020zio,Perivolaropoulos:2021jda,Akarsu:2024qiq,DiValentino:2025sru}. Foremost among these is the persistent disagreement on the value of the Hubble constant, $H_0$ (see~\cite{Verde:2019ivm,DiValentino:2020zio,DiValentino:2021izs,Perivolaropoulos:2021jda,Shah:2021onj,Abdalla:2022yfr,DiValentino:2022fjm,Kamionkowski:2022pkx,Verde:2023lmm,Perivolaropoulos:2024yxv,DiValentino:2025sru} and references therein for a thorough overview). Specifically, the $H_0$ values extracted from CMB measurements~\cite{Planck:2018vyg,SPT-3G:2022hvq,ACT:2025fju} under the assumption of $\Lambda$CDM differ markedly from those obtained via local astrophysical determinations based on different calibration methods~\cite{Freedman:2020dne,Birrer:2020tax,Wu:2021jyk,Anderson:2023aga,Scolnic:2023mrv,Jones:2022mvo,Anand:2021sum,Freedman:2021ahq,Uddin:2023iob,Huang:2023frr,Pesce:2020xfe,Kourkchi:2020iyz,Schombert:2020pxm,Blakeslee:2021rqi,deJaeger:2022lit,Murakami:2023xuy,Breuval:2024lsv,Freedman:2024eph,Riess:2024vfa,Vogl:2024bum,Gao:2024kkx,Scolnic:2024hbh,Said:2024pwm,Boubel:2024cqw,Scolnic:2024oth,Li:2025ife,Jensen:2025aai}. Known as the $H_0$ tension, this disagreement now exceeds the $5\sigma$ level~\cite{Riess:2022shoes,Murakami:2023xuy,Breuval:2024lsv} and represents one of the most urgent open problems facing $\Lambda$CDM. Efforts to address it have proceeded along two complementary fronts: scrutiny of potential systematic effects~\cite{Efstathiou:2020wxn,Mortsell:2021nzg,Mortsell:2021tcx,Riess:2022shoes,Sharon:2023ioz,Murakami:2023xuy,Riess:2023bfx,Bhardwaj:2023mau,Brout:2023wol,Dwomoh:2023bro,Uddin:2023iob,Riess:2024ohe,Freedman:2024eph,Riess:2024vfa,Colgain:2026ryf,Colgain:2024mtg}, and the development of alternative cosmological scenarios~\cite{Knox:2019rjx,Jedamzik:2020zmd,DiValentino:2021izs,Perivolaropoulos:2021jda,Schoneberg:2021qvd,Abdalla:2022yfr,Kamionkowski:2022pkx,Khalife:2023qbu,Hu:2023jqc,DiValentino:2025sru,Vagnozzi:2023nrq}. Beyond $H_0$, the so-called $S_8$ tension has also drawn attention~\cite{Nunes:2021ipq,Abdalla:2022yfr,DiValentino:2020vvd,DiValentino:2018gcu} (see also~\cite{Adil:2023jtu,Akarsu:2024hsu}), referring to the mismatch between the matter clustering amplitude predicted from early-universe data such as Planck CMB and the values recovered at late times via weak lensing and galaxy clustering, although its statistical relevance appears to have weakened in light of the recent KiDS-Legacy release~\cite{DES:2021bvc,DES:2021vln,KiDS:2020suj,Asgari:2019fkq,Joudaki:2019pmv,DAmico:2019fhj,Kilo-DegreeSurvey:2023gfr,Troster:2019ean,Heymans:2020gsg,Dalal:2023olq,Chen:2024vvk,ACT:2024okh,Sailer:2024coh,Harnois-Deraps:2024ucb,Dvornik:2022xap,DES:2021wwk,Wright:2025xka}. On top of these, the latest BAO analyses from the DESI collaboration, when combined with other leading cosmological probes, point toward dynamical dark energy (DDE)~\cite{DESI:2024mwx,DESI:2024aqx,DESI:2024jis,DESI:2025zgx,DESI:2025fii}. The strength of this preference depends on the dark-energy parametrization, the CMB likelihood, and the supernova compilation. Related trends have also been reported in combinations that omit either CMB information or DESI BAO~\cite{DES:2024jxu,DES:2025bxy}. These findings motivate tests of specific alternatives to $\Lambda$CDM rather than a model-independent claim of evolving dark energy. A wide variety of models have been put forward to alleviate the prevailing cosmological tensions~\cite{DiValentino:2025sru}, and a sizable fraction of them include a DDE ingredient that plays a non-trivial role at late and/or early cosmic times~\cite{DESI:2024mwx,Cortes:2024lgw,Shlivko:2024llw,Luongo:2024fww,Yin:2024hba,Gialamas:2024lyw,Dinda:2024kjf,Najafi:2024qzm,Wang:2024dka,Tada:2024znt,Carloni:2024zpl,Chan-GyungPark:2024mlx,DESI:2024kob,Bhattacharya:2024hep,Ramadan:2024kmn,Bhattacharya:2024kxp,Orchard:2024bve,Hernandez-Almada:2024ost,Pourojaghi:2024tmw,Giare:2024gpk,Reboucas:2024smm,Giare:2024ocw,Chan-GyungPark:2024brx,Menci:2024hop,Li:2024qus,Li:2024hrv,Notari:2024zmi,Gao:2024ily,Fikri:2024klc,Jiang:2024xnu,Zheng:2024qzi,Gomez-Valent:2024ejh,RoyChoudhury:2024wri,RoyChoudhury:2025dhe,Lewis:2024cqj,Wolf:2025jlc,Shajib:2025tpd,Giare:2025pzu,Chaussidon:2025npr,Kessler:2025kju,Alvarez:2025kma}.

The most basic phenomenological realization of DDE replaces the cosmological constant by a dark energy fluid characterized by an effective equation of state (EoS) parameter $w$ whose constant value is allowed to depart from $-1$, that is, $w_\mathrm{DE} \neq -1$; this scenario is commonly known as the $w$CDM model. A natural extension consists of letting the EoS parameter evolve with time, the canonical example being the Chevallier-Polarski-Linder (CPL) parametrization~\cite{Chevallier:2000qy, Linder:2002et}, where $w_\mathrm{DE}$ takes the form of a linear function of the scale factor in the Robertson--Walker (RW) metric. The CPL parametrization can cross the phantom divide, $w_\mathrm{DE}=-1$, for suitable parameter values, whereas a constant-$w$ model cannot cross it dynamically. That said, describing DDE purely through its EoS parameter is insufficient when modeling scenarios in which DDE develops negative effective energy densities at earlier epochs, since the corresponding EoS parameter can diverge when the effective density passes through zero~\cite{Ozulker:2022slu}.

An alternative route to explaining cosmic acceleration is to modify the gravitational sector itself, rather than attributing the observed dynamics entirely to a phenomenological dark energy fluid within GR. One promising example is Schouten--Codazzi Gravity (SCG)~\cite{Sussman:2024mzs}, a second-order theory derived from the Codazzi differential condition imposed on the Schouten tensor. Unlike its predecessor, Cotton Gravity~\cite{Harada:2021cotton, Mantica:2022flg}, SCG remains well defined in conformally flat spacetimes and is therefore naturally applicable to cosmology. In a spatially flat FLRW geometry, SCG modifies the Friedmann equation through two integration constants: $\lambda_0$, which acts as a cosmological constant, and $\gamma_0$, which generates an additional effective energy-density contribution scaling as $(1+z)$. The latter behaves as a geometric fluid with the fixed equation of state $w_\mathrm{eff}=-2/3$, providing a theoretically motivated extension of $\Lambda$CDM that reduces exactly to the standard model when $\gamma_0=0$. 

The appearance of an effective component with $w_{\mathrm{eff}}=-2/3$ is not unique to SCG and has independent motivation from both symmetry considerations and topological-defect cosmology. Since a separately conserved component with a constant equation of state evolves as $\rho\propto(1+z)^{3(1+w)}$, the SCG contribution proportional to $(1+z)$ has precisely the background scaling associated with $w=-2/3$. Ref.~\cite{Vazquez:2012ag} introduced a component with this equation of state as ``missing matter'' in a two-component dark-energy model motivated by a conformal-time symmetry relating the big bang to the future conformal singularity. Although the observational analysis presented therein was consistent with $\Lambda$CDM, it did not exclude a nonzero contribution from this additional component. The same equation of state arises from a frustrated or effectively static network of cosmological domain walls. Ref.~\cite{Conversi:2004pi} showed that the viability of domain walls as a dark-energy component depends sensitively on the adopted cosmological priors, while the nonparametric reconstruction of the $O_m$ diagnostic presented in Ref.~\cite{Escamilla-Rivera:2015odt} found a preference for an effective component with $w=-2/3$. More recently, Ref.~\cite{An:2025vfz} investigated whether topological defects could mimic the dynamical dark-energy behavior suggested by DESI observations and found that a percent-level domain-wall contribution can modestly improve the fit relative to $\Lambda$CDM. These results provide additional phenomenological motivation for studying the $(1+z)$ contribution generated geometrically in SCG. Nevertheless, this correspondence holds only at the homogeneous-background level: the SCG geometric sector and a physical domain-wall network need not possess the same perturbations, anisotropic stress, or clustering properties.

In this work, we perform a background-level observational test of SCG using cosmological observations. Our analysis incorporates three alternative Type~Ia supernova compilations--Union3~\cite{Rubin:2023jdq}, DES-SN5YR~\cite{Abbott:2024des}, and Pantheon$+$~\cite{Brout:2022panth,Scolnic:2021amr} (with a low-redshift cut $z > 0.01$ and the absolute magnitude $M$ treated as a nuisance parameter)---alongside DESI~DR2 BAO measurements~\cite{DESI:2025dr2}, a compressed CMB likelihood, and a BBN prior on the baryon density. We employ Bayesian inference via \texttt{UltraNest}~\cite{Buchner:2021ultranest} and \texttt{emcee}~\cite{ForemanMackey:2013emcee} to explore 15 dataset combinations to constrain $\gamma_0$ and compare the resulting background expansion with that of $\Lambda$CDM.

The paper is organized as follows: Section~\ref{sec:scg} details the SCG framework; Section~\ref{sec:data} describes the datasets and methodology; Section~\ref{sec:results} presents our results; and Section~\ref{sec:conclusions} summarizes our findings.
\section{Schouten-Codazzi Gravity}\label{sec:scg}

Schouten--Codazzi gravity (SCG) is a second-order phenomenological gravity theory, motivated by the geometric relation between the Schouten and Cotton tensors \cite{Sussman:2024mzs}. In four dimensions, the Cotton tensor is obtained as a Codazzi constraint of the Schouten tensor. This observation underlies the formulation of Cotton gravity \cite{Harada:2021cotton, Mantica:2022flg}, a third-order gravity theory. The Cotton tensor vanishes for conformally flat geometries, as in the Friedmann-Lema\^itre-Robertson-Walker (FLRW) metric, which limits the applicability of Cotton gravity to Cosmology. 

SCG follows a different construction. Rather than imposing the differential Codazzi condition on the Schouten tensor, it introduces a second-rank symmetric tensor which satisfies this condition. The resulting field equations remain second order in the metric. The algebraic form of this tensor, referred to as the prospective Codazzi tensor, is prescribed using the invariant algebraic structure and symmetries of the spacetime. Since the present work is concerned with cosmological observations, we will restrict our discussion of SCG to the necessary ingredients to determine the dynamics of the FLRW spacetimes. 

\subsection{Field equations}

In SCG, we consider a four-dimensional Lorentzian spacetime with metric signature ($-,+,+,+$) and the Levi-Civita connection. We work in geometrical units $c=1$ and define $\kappa= 8\pi G$. Following the SCG convention adopted in \cite{Sussman:2024mzs}, the Schouten tensor is defined as 
\begin{equation}
S_{ab}=R_{ab}-\frac16 R g_{ab}.
\end{equation}

In general relativity without a cosmological constant, the vacuum field equations can be written as $S_{ab}=0$, while in the presence of matter $S_{ab}=\kappa \hat{T}_{ab}$, where $\hat{T}_{ab} = T_{ab}-\frac13 Tg_{ab} $, and $T_{ab}$ is the energy-momentum tensor. By adding the prospective Codazzi tensor $F_{ab}$, the SCG field equations take the form 
\begin{equation}
S_{ab}+F_{ab}=\kappa \hat{T}_{ab},\label{eq:scgfield}
\end{equation}
where $F_{ab}$ is a symmetric tensor that satisfies the Codazzi differential condition
\begin{equation}
    \mathcal{F}_{abc} \equiv \nabla_a F_{bc}
    - \nabla_c F_{ba} = 0.
    \label{eq:codazzi}
\end{equation}
The key feature of SCG is that $F_{ab}$ is
\emph{prescribed} from the algebraic structure of the spacetime before the field equations are solved, rather than being determined by them. This removes the underdeterminacy that renders Cotton Gravity~\cite{Harada:2021cotton} ill-defined in conformally flat spacetimes, while keeping the theory second-order in the metric~\cite{Sussman:2024mzs}.

The SCG field equations can also be written in an effective Einstein form 
\begin{equation}
G_{ab} + F_{ab} -F g_{ab} = \kappa T_{ab},  \label{eq:scg_field}
\end{equation}
where $F=g^{cd}F_{cd}$. This representation facilitates comparison with the Einstein field equations.

\subsection{FLRW solution}
 
We consider a spatially flat Friedmann--Lema\^{\i}tre--Robertson--Walker (FLRW) spacetime with line element
\begin{equation}
    ds^2 = -dt^2 + a^2(t) \; (dx^2+dy^2+dz^2), \label{eq:flrw_metric}
\end{equation}
where $a(t)$ is the scale factor and $(t,x,y,z)$ are comoving coordinates.
 
In this spacetime, both the Ricci tensor and a perfect-fluid energy-momentum tensor belong to Segr\'{e} type $[1,(1,1,1)]$: one timelike and three equal spacelike eigenvalues. For consistency in Eq.~(\ref{eq:scg_field}), $\mathcal{F}_{ab}$ must share the same algebraic type, fixing the ansatz
\begin{equation}
\tensor{F}{^a _b} = -F_{0}(t) \, u^a u_b + F(t) \, (x^a x_b + y^a y_b + z^a z_b),  \label{eq:Fab_ansatz}
\end{equation}
where $u^a$ is the comoving four-velocity and $\{u^a, x^a, y^a, z^a\}$ is the orthonormal tetrad associated with the comoving coordinates $(t,x,y,z)$. Imposing the Codazzi condition Eq.~(\ref{eq:codazzi}) on our ansatz Eq.~(\ref{eq:Fab_ansatz}), we adopt the cosmological branch considered in Ref.~\cite{Sussman:2024mzs},
\begin{equation}
    F_0 = \lambda_0, \qquad F=\lambda_0 + \frac{\gamma_0}{a},\label{eq:F_solution}
\end{equation}
where $\lambda_0$ and $\gamma_0$ are integration constants. 

Substitution of Eq.~(\ref{eq:F_solution}) into the field equations Eq.~(\ref{eq:scg_field}) produces the modified Friedmann equations 
\begin{align}
    H^2 &= \frac{\kappa}{3}\rho - \frac{k}{a^2}+\lambda_0+ \frac{\gamma_0}{a},\label{eq:friedmann}\\
    \frac{\ddot{a}}{a} &= -\frac{\kappa}{6} (\rho +3p) + \frac{\gamma_0}{2a}+ \lambda_0.\label{eq:raychaudhuri}
\end{align}

The background energy-momentum conservation equation 
\begin{equation}
    \dot{\rho} = -3(\rho+p)\frac{\dot{a}}{a},\label{eq:conservation}
\end{equation}
is unaffected, so minimally coupled matter species obey the usual background continuity equations. 

\subsection{Cosmological parameters and the \texorpdfstring{$\Lambda$}{Lambda}CDM limit}

Defining dimensionless density parameters $\Omega_{\lambda_0} \equiv \lambda_0/H_0^2$ and $\Omega_{\gamma_0} \equiv \gamma_0/H_0^2$ for the two SCG integration constants, dividing \cref{eq:friedmann} by $H_0^2$ gives the dimensionless expansion rate
\begin{equation}
    E^2(z) \equiv \frac{H^2}{H_0^2}= \Omega_m(1+z)^3 + \Omega_r(1+z)^4+ \Omega_{\lambda_0} +\Omega_{\gamma_0}(1+z),\label{eq:Ez}
\end{equation}
with the flatness constraint
\begin{equation}
    \Omega_m + \Omega_r + \Omega_{\lambda_0} + \Omega_{\gamma_0} = 1.\label{eq:flatness}
\end{equation}
Because of this constraint, $\Omega_{\lambda_0}$ is a derived quantity. The independent cosmological parameters of SCG that we vary in the inference are $\{H_0,\,\Omega_\mathrm{cdm},\,\Omega_b,\,\Omega_{\gamma_0}\}$, where $H_0$ is the present-day Hubble constant, $\Omega_\mathrm{cdm}$ and $\Omega_b$ are the present-day cold dark matter and baryon density parameters (so that the total matter density appearing in~\cref{eq:flatness} is $\Omega_m \equiv \Omega_\mathrm{cdm} + \Omega_b$), and $\Omega_{\gamma_0}$ is the dimensionless counterpart of the SCG integration constant $\gamma_0$ defined above. The radiation density is calculated from the adopted standard radiation sector; $\Omega_r$ consequently depends on $H_0$.

Setting $\gamma_0 = 0$,~\cref{eq:friedmann} becomes the standard $\Lambda$CDM Friedmann equation with $\Lambda = 3\lambda_0$. In this limit $\Omega_{\lambda_0}$ coincides with the $\Lambda$CDM dark energy density parameter $\Omega_\Lambda \equiv \Lambda/(3 H_0^2)$, but the two quantities belong to different models: throughout this paper we reserve the lowercase subscript $\Omega_{\lambda_0}$ exclusively for the SCG parameter associated with $\lambda_0$, and the uppercase $\Omega_\Lambda$ exclusively for the $\Lambda$CDM cosmological-constant density parameter. The parameter $\gamma_0$ is therefore the sole departure of SCG from $\Lambda$CDM at the level of the cosmological background.

The effective energy density associated with $\gamma_0$ scales as $\rho_{\gamma_0} \propto a^{-1}$, corresponding to the rigid geometric equation of state
\begin{equation}
    w_\mathrm{eff} = -\tfrac{2}{3}.
    \label{eq:weff}
\end{equation}
The value $w_{\gamma_0}=-2/3$ is fixed for this background component; it is not a free equation-of-state parameter for the combined effective dark sector. At fixed $H_0$, $\Omega_m$, and $\Omega_r$, enforcing flatness gives $E^2_{\rm SCG}(z)-E^2_{\Lambda{\rm CDM}}(z)=\Omega_{\gamma_0}z$. Thus positive $\Omega_{\gamma_0}$ increases the expansion rate at $z>0$, while negative values decrease it wherever $E^2>0$. Although this contribution grows with redshift, its ratio to the matter term decreases as $(1+z)^{-2}$ and its ratio to the radiation term as $(1+z)^{-3}$. Its suppression at early times motivates a standard early-universe approximation at the background level, but does not establish equivalence of the perturbations.

\section{Data and Statistical Methodology}\label{sec:data}
 Our analysis combines four cosmological probes: three independent Type Ia supernova (SNe~Ia) compilations (DES-SN5YR~\cite{DES:2025bxy}, Union3~\cite{Rubin:2023jdq}, and Pantheon$+$~\cite{Brout:2021mpj}, used one at a time as the SNe~Ia channel), Baryon Acoustic Oscillation (BAO) measurements from DESI~DR2~\cite{DESI:2025zgx}, a compressed Cosmic Microwave Background (CMB) likelihood, and a Big Bang nucleosynthesis (BBN) prior on the baryon density when CMB measurements are not used. Table~\ref{tab:data_summary} summarizes the number of likelihood data points contributed by each dataset. 
 
\begin{table}[!htbp]
    \centering
    \caption{Summary of datasets used in this work and the number of likelihood data points contributed by each.}
    \label{tab:data_summary}
    \begin{ruledtabular}
    \begin{tabular}{lcc}
    Dataset & Data points & Redshift range \\ \hline
    Pantheon+ (SNe~Ia) & 1590 & $0.01 \leq z \leq 2.26$ \\
    DES-SN5YR (SNe~Ia)   & 1829 & $0.1\phantom{0} \leq z \leq 1.3\phantom{0}$ \\
    Union3 (SNe~Ia)      & 22   & $0.05 \leq z \leq 1.0\phantom{0}$ \\
    DESI~DR2 (BAO)       & 13   & $0.295 \leq z \leq 2.33$ \\
    Compressed CMB       & 3    & $z_\star = 1089$ \\
    BBN prior on $\omega_b$ & 1 & --- \\
    \end{tabular}
    \end{ruledtabular}
\end{table}

 \subsection{Statistical methodology}

Parameter estimation and Bayesian model comparison were performed using two different samplers. Posterior samples for parameter estimation were generated with the \texttt{emcee} package \cite{ForemanMackey:2013emcee} using standard values as an ensemble MCMC sampler, while Bayesian evidence required for model comparison between the $\Lambda$CDM and SCG models was computed with the nested sampler \texttt{UltraNest} \cite{Buchner:2021ultranest}.

Parameter inference is based on the Bayesian posterior
\begin{equation}
    P(\bm{\theta}\mid\mathbf{d}) \propto
    \mathcal{L}(\mathbf{d}\mid\bm{\theta})\,\pi(\bm{\theta}),
    \label{eq:bayes}
\end{equation}
For Gaussian data, $\ln\mathcal{L}=-\chi^2/2+\mathrm{const}$. Likelihood normalizations must be treated consistently when comparing evidences for the same dataset. The free parameters are $\bm{\theta} = \{H_0,\,\Omega_\mathrm{cdm},\,\Omega_b,\,\Omega_{\gamma_0}\}$, with $\Omega_{\lambda_0}$ derived from the flatness constraint~\cref{eq:flatness}. For Pantheon$+$ fits the absolute magnitude $M$ is included as an additional nuisance parameter. All parameters are assigned uniform priors within the ranges listed in Table~\ref{tab:priors}.

\begin{table}[t]
\centering
\caption{Uniform prior ranges for the free parameters.}
\label{tab:priors}
\begin{tabular}{lcc}
\toprule
Parameter & Min & Max \\
\midrule
$H_0~(\mathrm{km\,s^{-1}\,Mpc^{-1}})$ & 60 & 90 \\
$\Omega_\mathrm{cdm}$                   & 0.10 & 0.50 \\
$\Omega_b$                              & 0.020 & 0.060 \\
$\Omega_{\gamma_0}$                     & $-0.50$ & $0.50$ \\
\bottomrule
\end{tabular}
\end{table}

The total log-likelihood sums over the active probes,
\begin{equation}
    \chi^2 = \sum_i \chi^2_i.
    \label{eq:chi2_total}
\end{equation}
For Type~Ia supernovae the chi-squared takes the multivariate Gaussian form
\begin{equation}
    \chi^2_\mathrm{SN} =
    \mathbf{r}^\top \mathbf{C}_\mathrm{SN}^{-1}\,\mathbf{r},
    \qquad \mathbf{r} = \bm{\mu}_\mathrm{obs} - \bm{\mu}_\mathrm{th},
    \label{eq:chi2_sn}
\end{equation}
where $\mathbf{C}_\mathrm{SN}$ is the covariance matrix, including statictical and systematic contributions, supplied with each compilation, and $\mu_{\rm th}$ is the theoretical distance modulus given by
\begin{equation}
    \mu_\mathrm{th}(z) =
    5\log_{10}\!\left[\frac{(1+z)\,D_M(z)}{\mathrm{Mpc}}\right] + 25,
    \label{eq:mu_th}
\end{equation}
with the comoving distance
\begin{equation}
    D_M(z) = \frac{c}{H_0}\int_0^z \frac{dz'}{E(z')},
    \label{eq:DM}
\end{equation}
using $E(z)$ from~\cref{eq:Ez}. For BAO,
\begin{equation}
    \chi^2_\mathrm{BAO} =
    \Delta\mathbf{d}^\top\,\mathbf{C}_\mathrm{BAO}^{-1}\,\Delta\mathbf{d},
    \label{eq:chi2_bao}
\end{equation}
where $\Delta\mathbf{d}$ is the 13 element vector of residuals in $D_V/r_d$, $D_M/r_d$, and $D_H/r_d$, and $\mathbf{C}_\mathrm{BAO}$ is the full $13\times13$ covariance matrix from~\cite{DESI:2025dr2}. The compressed CMB likelihood uses the same form with a $3\times3$ covariance matrix in $[\theta_*,\,\omega_b,\,\omega_m]$. The BBN prior contributes a Gaussian term
\begin{equation}
    \chi^2_\mathrm{BBN}
    = \left(\frac{\omega_b - 0.02196}{0.00063}\right)^2.
    \label{eq:chi2_bbn}
\end{equation}

We analyze 15 dataset combinations: three SN-only fits, three SN+BAO, three SN+CMB, three SN+BAO+CMB, plus BAO-only, CMB-only, and BAO+CMB configurations. Reported constraints are the posterior median with 68\% credible intervals.

For the evidence estimation we use the bayesian evidence defined as,
\begin{equation}
    \ln\mathcal{Z} = \ln\!\int \mathcal{L}(\bm{\theta})\,
    \pi(\bm{\theta})\,d\bm{\theta},
    \label{eq:lnZ}
\end{equation}
This enables model comparison via the log Bayes factor
\begin{equation}
    \ln B = \ln\mathcal{Z}_\mathrm{SCG}
          - \ln\mathcal{Z}_{\Lambda\mathrm{CDM}}.
    \label{eq:lnB}
\end{equation}

\section{Results}\label{sec:results}

\begin{figure*}[!t]
    \centering
    \includegraphics[width=\textwidth]{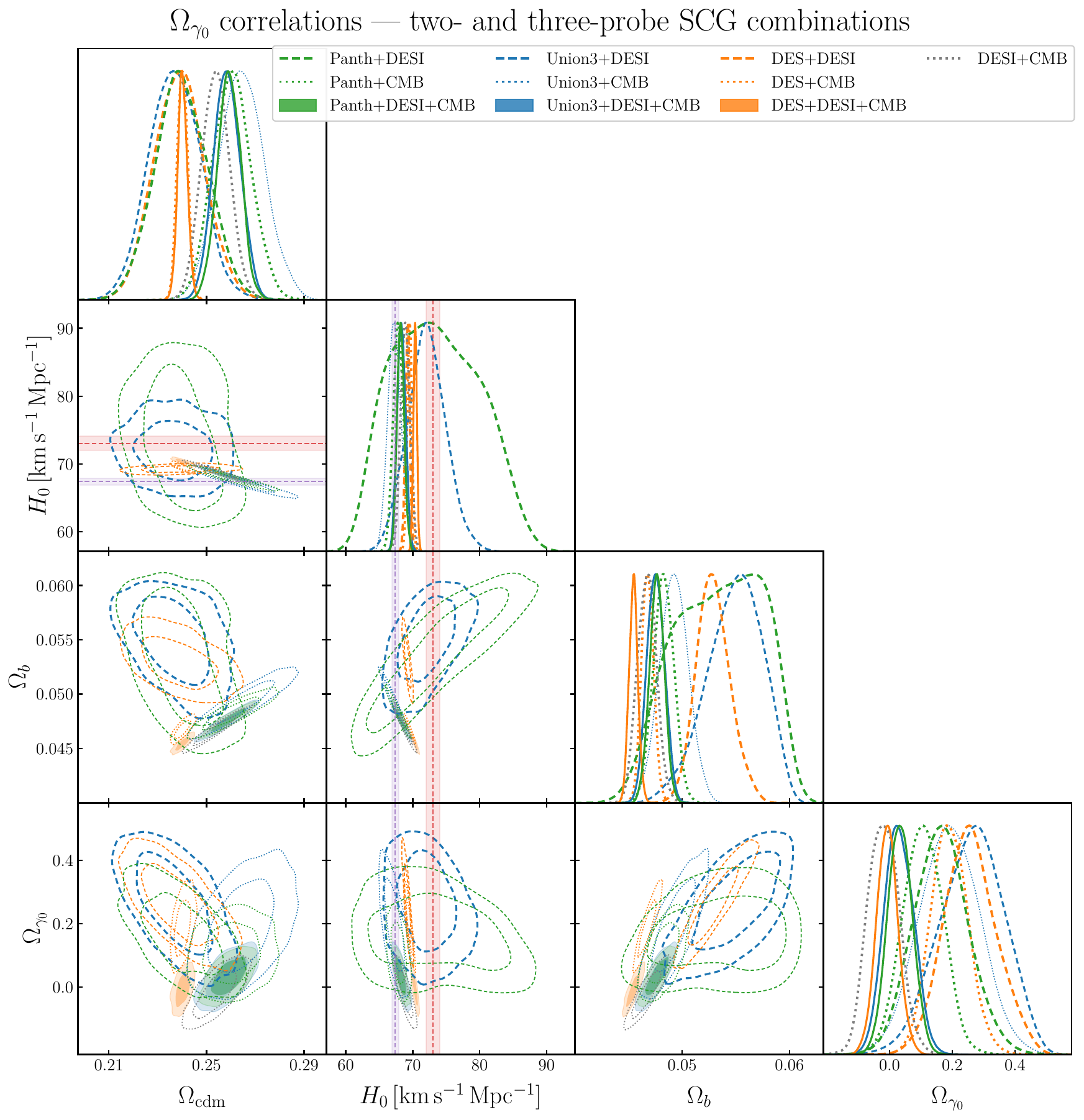}
    \caption{Marginalized one-dimensional posterior distributions and joint two-dimensional constraints on $\Omega_{\rm cdm}$, $H_0$, $\Omega_b$, and $\Omega_{\gamma0}$ for the two- and three-probe SCG combinations. Dashed curves and contours show the SNe~Ia$+$DESI combinations, while dotted curves and contours show the SNe~Ia$+$CMB combinations. The gray dotted curves correspond to DESI$+$CMB. Filled contours and solid one-dimensional distributions represent the full SNe~Ia$+$DESI$+$CMB combinations. Blue, orange, and green denote results obtained using Union3, DES-SN5YR, and Pantheon+, respectively. The inner and outer contours enclose the 68\% and 95\% credible regions. The purple and red shaded bands mark the 68\% intervals of the Planck~2018~\cite{Planck:2018vyg} and SH0ES~\cite{Riess:2022shoes} measurements of $H_0$, respectively.}
    \label{fig:corner}
\end{figure*}

Figure~\ref{fig:corner} illustrates how the different probe combinations constrain the SCG parameter $\Omega_{\gamma0}$ and its correlations with $\Omega_{\rm cdm}$, $H_0$, and $\Omega_b$. The two-probe combinations containing SNe~Ia favor positive values of $\Omega_{\gamma0}$, although the posterior position and width depend on both the supernova compilation and the complementary probe. The figure also shows that the strength and orientation of the parameter correlations change with the choice of probes, reflecting the different epochs and physical scales sampled by the data. The late-time SNe~Ia$+$DESI combinations produce relatively broad and, in some cases, curved degeneracy directions because both probes primarily constrain the recent expansion history. When compressed CMB information is included, the acoustic-scale and physical-density constraints rotate and narrow several of these degeneracies. This change is particularly visible in the $H_0$--$\Omega_b$ plane, where the correlation direction differs between the DESI- and CMB-based combinations, and in the $\Omega_{\gamma0}$--$H_0$ and $\Omega_{\gamma0}$--$\Omega_{\rm cdm}$ planes, where the addition of early-time information produces substantially tighter, more strongly oriented contours. The dependence of the contours on the supernova compilation further indicates that the inferred correlations are sensitive to the detailed late-time distance calibration rather than being determined by one probe alone.

Adding the third probe breaks much of these degeneracies and contracts the posterior contours around $\Omega_{\gamma0}=0$. The resulting constraints, $\Omega_{\gamma0}=0.029^{+0.048}_{-0.045}$, $-0.007^{+0.033}_{-0.034}$, and $0.033\pm0.039$ for the Union3-, DES-SN5YR-, and Pantheon+-based combinations, respectively, are all consistent with $\Lambda$CDM within $1\sigma$. The three-probe constraints on $H_0$ lie between the Planck and SH0ES determinations but remain closer to the lower values, confirming that the additional SCG contribution does not by itself fully resolve the $H_0$ tension.

Further, we present constraints on $\Omega_{\gamma_0}$ from all the 15 dataset combinations below. All quoted values are posterior medians with 68\% credible intervals obtained with \texttt{emcee}. We compare these constraints with $\Lambda$CDM fits using the same dataset combinations.

\subsection{Single- and two-probe constraints on \texorpdfstring{$\Omega_{\gamma_0}$}{gamma0}}
\label{sec:results:2probe}

With a single dataset, the posterior on $\Omega_{\gamma_0}$ remains broad, with $1\sigma$ widths between $\sim 0.13$ and $\sim 0.32$. None of the single-probe configurations resolves $\Omega_{\gamma_0}$ away from zero at more than $1\sigma$, indicating that the individual datasets leave substantial degeneracies with the other background parameters.
Combining each SNe~Ia compilation with BAO or compressed CMB information shifts the posterior toward positive $\Omega_{\gamma_0}$. The values in parentheses below summarize approximate Gaussian departures from zero, rather than calibrated detection significances:
\begin{align}
    \Omega_{\gamma_0}^{\rm Union3+DESI}  &= +0.267^{+0.097}_{-0.109} && (2.6\sigma), \nonumber \\
    \Omega_{\gamma_0}^{\rm Union3+CMB}   &= +0.191^{+0.097}_{-0.096} && (2.0\sigma), \nonumber \\
    \Omega_{\gamma_0}^{\rm DES+DESI}     &= +0.257^{+0.086}_{-0.082} && (3.1\sigma), \nonumber \\
    \Omega_{\gamma_0}^{\rm DES+CMB}      &= +0.189^{+0.061}_{-0.057} && (3.2\sigma), \nonumber \\
    \Omega_{\gamma_0}^{\rm Panth+DESI}   &= +0.164^{+0.086}_{-0.087} && (1.9\sigma), \nonumber \\
    \Omega_{\gamma_0}^{\rm Panth+CMB}    &= +0.107^{+0.063}_{-0.065} && (1.7\sigma).
\end{align}
The direction of the shift is consistent across the three SNe~Ia compilations and both complementary probes:  all six SNe~Ia$+$\{BAO,CMB\} combinations push $\Omega_{\gamma_0}$ toward positive values, with the largest approximate departures from zero obtained for DES-SN5YR. In contrast, the BAO$+$CMB combination---which contains no SNe~Ia information---returns $\Omega_{\gamma_0}^{\rm DESI+CMB} = -0.020 \pm 0.049$ ($0.4\sigma$), fully consistent with $\Lambda$CDM.

\subsection{Three-probe joint constraints}
\label{sec:results:3probe}

The joint constraints follow from the three full SNe~Ia$+$BAO$+$CMB combinations are summarized in Table~\ref{tab:results_3probe}.

Adding the third probe shifts the posterior of $\Omega_{\gamma_0}$ toward zero and reduces its uncertainty (Table~\ref{tab:results_3probe}). The three combinations yield
\begin{align}
    \Omega_{\gamma_0}^{\rm Union3+DESI+CMB}  &= +0.029^{+0.048}_{-0.045} && (0.6\sigma), \nonumber \\
    \Omega_{\gamma_0}^{\rm DES+DESI+CMB}     &= -0.007^{+0.033}_{-0.034} && (0.2\sigma), \nonumber \\
    \Omega_{\gamma_0}^{\rm Panth+DESI+CMB}   &= +0.033^{+0.039}_{-0.039} && (0.8\sigma),
\end{align}
These values are all consistent with the $\Lambda$CDM limit at the $1\sigma$ level. The marginal uncertainty decreases with respect to the corresponding two-probe constraints.

The matter sector is in mutual agreement between SCG and $\Lambda$CDM at the $1\sigma$ level, as the two-dimensional posteriors of Fig.~\ref{fig:corner} make explicit. SCG returns $\Omega_\mathrm{cdm} = 0.258$--$0.260$ for Union3 and Pantheon$+$, and $\Omega_\mathrm{cdm} = 0.240$ for DES-SN5YR; the corresponding $\Lambda$CDM fits give $0.256$, $0.257$ and $0.240$, respectively, with comparable uncertainties. The baryon density is essentially identical across models, $\Omega_b = 0.0455$--$0.0476$, with similar marginalized constraints in both models. 

\begin{figure}[!htbp]
    \centering
    \includegraphics[width=\columnwidth]{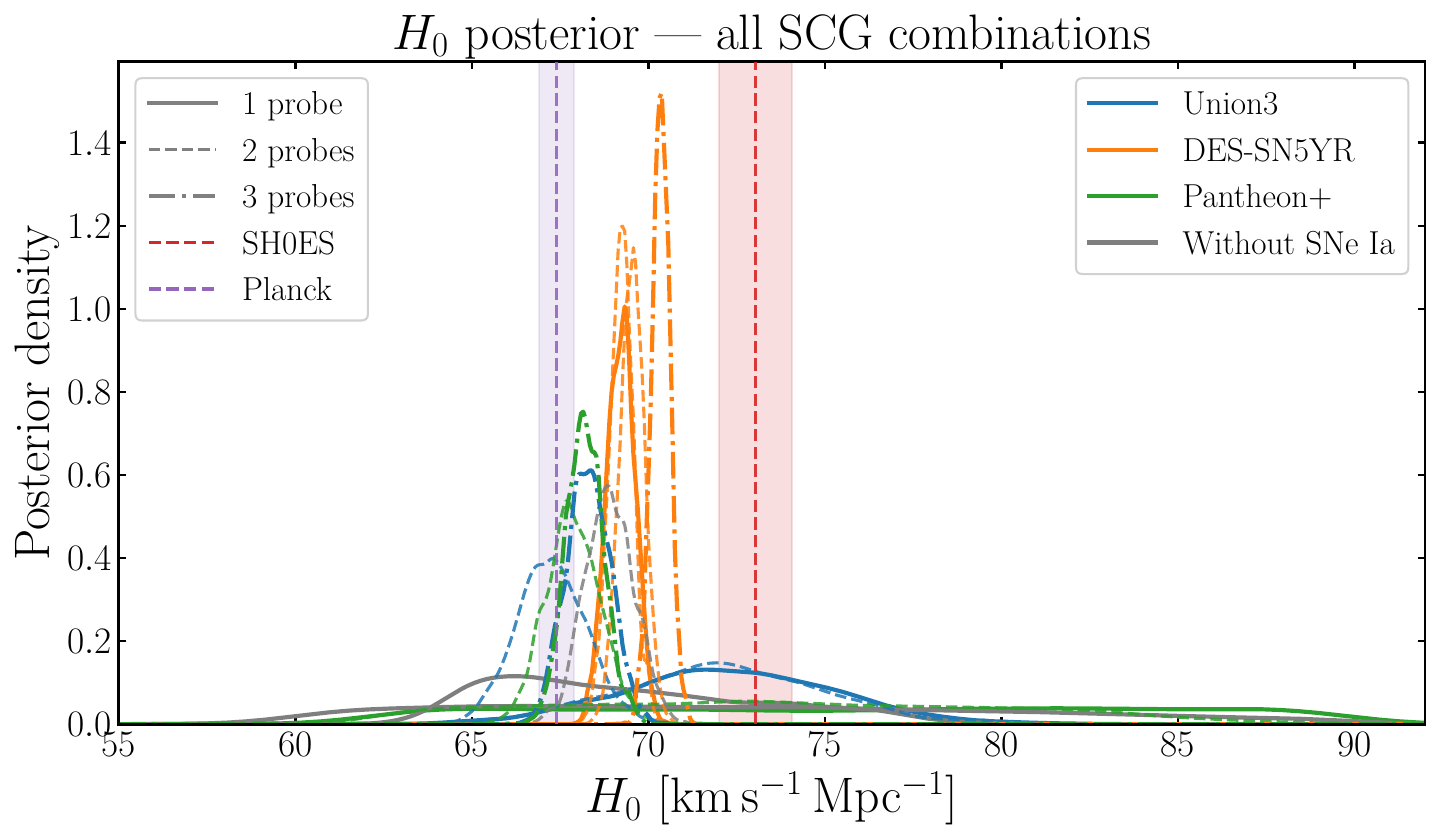}
    \caption{Marginalized posterior distributions of $H_0$ for all 15 SCG data combinations, grouped according to the number of probes employed. Solid, dashed, and dash-dotted curves denote the one-, two-, and three-probe combinations, respectively. Blue, orange, and green curves correspond to combinations involving Union3, DES-SN5YR, and Pantheon+, while gray curves denote combinations without SNe~Ia. The purple and red dashed vertical lines indicate the Planck~2018~\cite{Planck:2018vyg} and SH0ES~\cite{Riess:2022shoes} central values, and the corresponding shaded regions show their 68\% confidence intervals.}
    \label{fig:h0_tension}
\end{figure}

Across the 15 data combinations shown in Fig.~\ref{fig:h0_tension}, the inferred value of $H_0$ depends primarily on the probes included and, in particular, on the adopted SNe~Ia compilation. The five individual constraints from Union3, DES-SN5YR, Pantheon+, DESI, and CMB are comparatively broad, reflecting the limited ability of a single probe to break the degeneracies of the SCG background. The seven two-probe combinations--Union3$+$DESI, Union3$+$CMB, DES-SN5YR$+$DESI, DES-SN5YR$+$CMB, Pantheon$+$$+$DESI, Pantheon$+$$+$CMB, and DESI$+$CMB--produce narrower posteriors whose peaks generally lie between the Planck and SH0ES reference values. In particular, the combinations involving DES-SN5YR tend to favor somewhat larger values of $H_0$ than the corresponding Union3 and Pantheon$+$ combinations, thereby shifting the posterior modestly toward the local SH0ES determination. The three-probe constraints preserve this dependence on the supernova sample, yielding $H_0=68.30\pm0.62~\mathrm{km\,s^{-1}\,Mpc^{-1}}$ for Union3$+$DESI$+$CMB, $70.35\pm0.26~\mathrm{km\,s^{-1}\,Mpc^{-1}}$ for DES-SN5YR$+$DESI$+$CMB, and $68.19\pm0.53~\mathrm{km\,s^{-1}\,Mpc^{-1}}$ for Pantheon$+$$+$DESI$+$CMB. The Union3- and Pantheon$+$-based results remain close to the Planck inference, whereas the DES-SN5YR combination peaks in the intermediate region between the Planck and SH0ES bands. Since the SCG and $\Lambda$CDM central values differ by at most $0.31~\mathrm{km\,s^{-1}\,Mpc^{-1}}$, SCG alone does not resolve the $H_0$ tension. Nevertheless, the intermediate peaks obtained for some two- and three-probe combinations indicate that the model can marginally reduce the discrepancy for particular choices of data, especially those involving DES-SN5YR.

\begin{table*}[!htbp]
\centering
\caption{Marginalized parameter constraints (posterior medians and 68\% credible intervals) for the three full SNe~Ia$+$BAO$+$CMB combinations, comparing SCG against $\Lambda$CDM. $\Omega_{\lambda_0}$ in SCG is derived via the flatness condition~\cref{eq:flatness}; $\Omega_\Lambda$ is its $\Lambda$CDM counterpart. The bottom row reports the tension of $\Omega_{\gamma_0}$ with the $\Lambda$CDM limit $\Omega_{\gamma_0} = 0$.}
\label{tab:results_3probe}
\begin{ruledtabular}
\begin{tabular}{lcccccc}
 & \multicolumn{2}{c}{Union3$+$DESI$+$CMB}
 & \multicolumn{2}{c}{DES-SN5YR$+$DESI$+$CMB}
 & \multicolumn{2}{c}{Pantheon$+$$+$DESI$+$CMB} \\
\cline{2-3}\cline{4-5}\cline{6-7}
 & SCG & $\Lambda$CDM & SCG & $\Lambda$CDM & SCG & $\Lambda$CDM \\
\hline
$H_0~[\mathrm{km\,s^{-1}\,Mpc^{-1}}]$
  & $68.30^{+0.64}_{-0.62}$ & $68.61^{+0.41}_{-0.42}$
  & $70.35^{+0.25}_{-0.26}$ & $70.31^{+0.16}_{-0.16}$
  & $68.19^{+0.53}_{-0.53}$ & $68.50^{+0.39}_{-0.34}$ \\
$\Omega_\mathrm{cdm}$
  & $0.2584^{+0.0055}_{-0.0054}$ & $0.2561^{+0.0043}_{-0.0043}$
  & $0.2401^{+0.0021}_{-0.0018}$ & $0.2404^{+0.0016}_{-0.0017}$
  & $0.2592^{+0.0050}_{-0.0047}$ & $0.2572^{+0.0037}_{-0.0041}$ \\
$\Omega_b$
  & $0.0475^{+0.0008}_{-0.0008}$ & $0.0471^{+0.0004}_{-0.0004}$
  & $0.0455^{+0.0004}_{-0.0004}$ & $0.0455^{+0.0002}_{-0.0002}$
  & $0.0476^{+0.0007}_{-0.0007}$ & $0.0472^{+0.0004}_{-0.0004}$ \\
$\Omega_{\gamma_0}$
  & $+0.029^{+0.048}_{-0.045}$ & ---
  & $-0.007^{+0.033}_{-0.034}$ & ---
  & $+0.033^{+0.039}_{-0.039}$ & --- \\
$\Omega_{\lambda_0}\;/\;\Omega_\Lambda$
  & $0.6650^{+0.0501}_{-0.0528}$ & $0.6967^{+0.0048}_{-0.0047}$
  & $0.7214^{+0.0350}_{-0.0347}$ & $0.7140^{+0.0019}_{-0.0018}$
  & $0.6596^{+0.0421}_{-0.0424}$ & $0.6955^{+0.0044}_{-0.0041}$ \\
$\ln\mathcal{Z}$
  & $-37.27 \pm 0.48$  & $-35.37 \pm 0.64$
  & $-862.73 \pm 0.41$ & $-860.18 \pm 0.75$
  & $-729.42 \pm 0.31$ & $-728.29 \pm 0.99$ \\
\hline
$|\Omega_{\gamma_0}/\sigma|$
  & \multicolumn{2}{c}{$0.6\sigma$}
  & \multicolumn{2}{c}{$0.2\sigma$}
  & \multicolumn{2}{c}{$0.8\sigma$} \\
\end{tabular}
\end{ruledtabular}
\end{table*}

\begin{figure}[!htbp]
    \centering
    \includegraphics[width=\columnwidth]{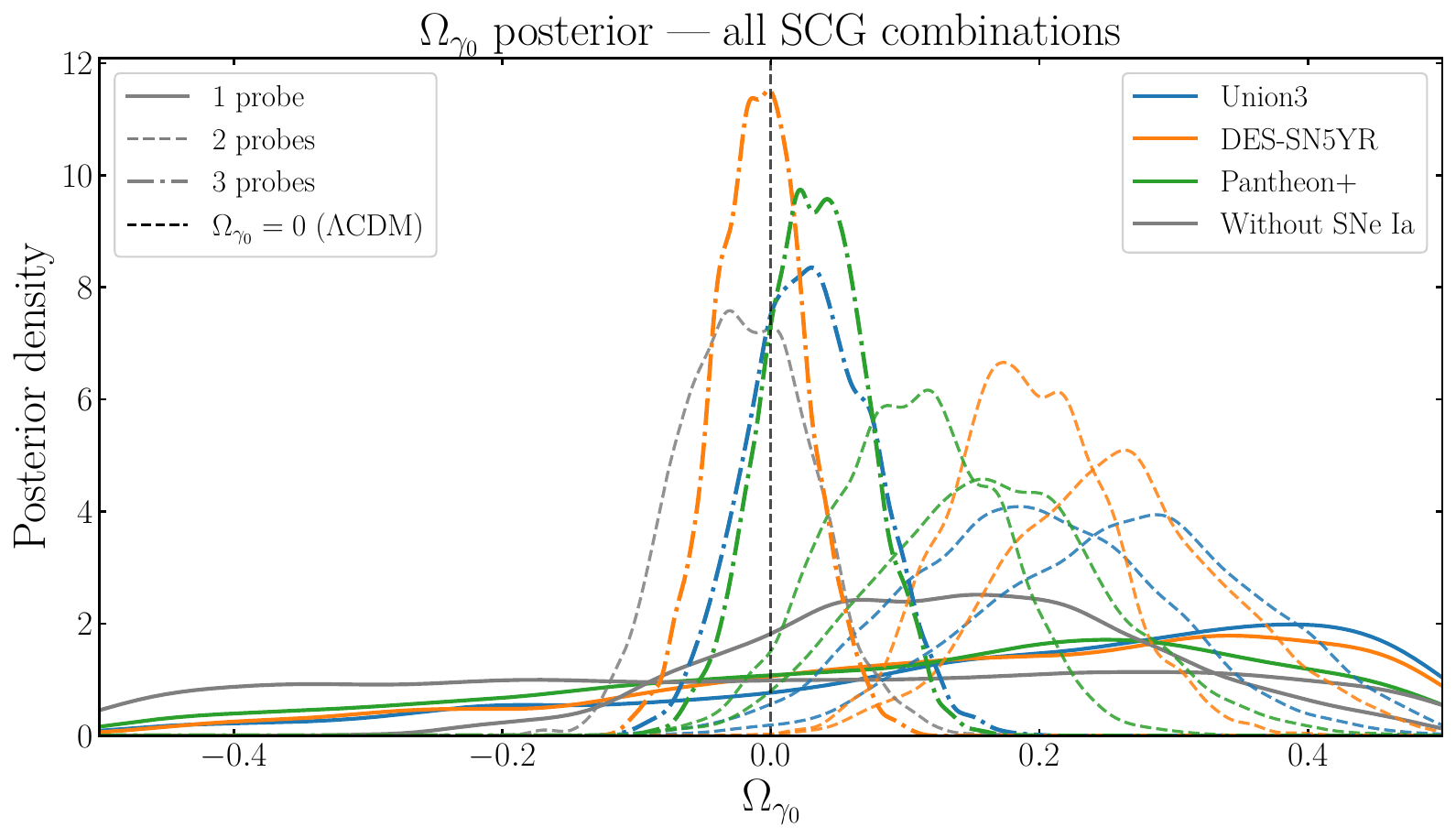}
    \caption{Marginalized posterior distributions of the additional SCG contribution $\Omega_{\gamma0}$ for all 15 data combinations, grouped according to the number of probes employed. Solid, dashed, and dash-dotted curves denote the one-, two-, and three-probe combinations, respectively. Blue, orange, and green curves correspond to combinations involving Union3, DES-SN5YR, and Pantheon+, while gray curves denote combinations without SNe~Ia. The vertical black dashed line marks the $\Lambda$CDM limit, $\Omega_{\gamma0}=0$.}
    \label{fig:tension_3probes}
\end{figure}

Figure~\ref{fig:tension_3probes} highlights how the marginalized posterior of $\Omega_{\gamma0}$ evolves as complementary information is added. The individual-probe posteriors are broad and, in some cases, visibly non-Gaussian, demonstrating that a single data set leaves substantial degeneracies between the SCG contribution and the remaining background parameters. The two-probe combinations involving SNe~Ia generally shift probability toward positive $\Omega_{\gamma0}$, with the position and shape of the posterior depending on both the supernova compilation and whether DESI or compressed CMB information is included. Some of these distributions exhibit extended tails indicating that the marginalized constraint is affected by curved or partially unresolved degeneracy directions. By contrast, the DESI$+$CMB posterior, which contains no SNe~Ia data, is concentrated around the $\Lambda$CDM limit.

\begin{figure}[!htbp]
    \centering
    \includegraphics[width=\columnwidth]{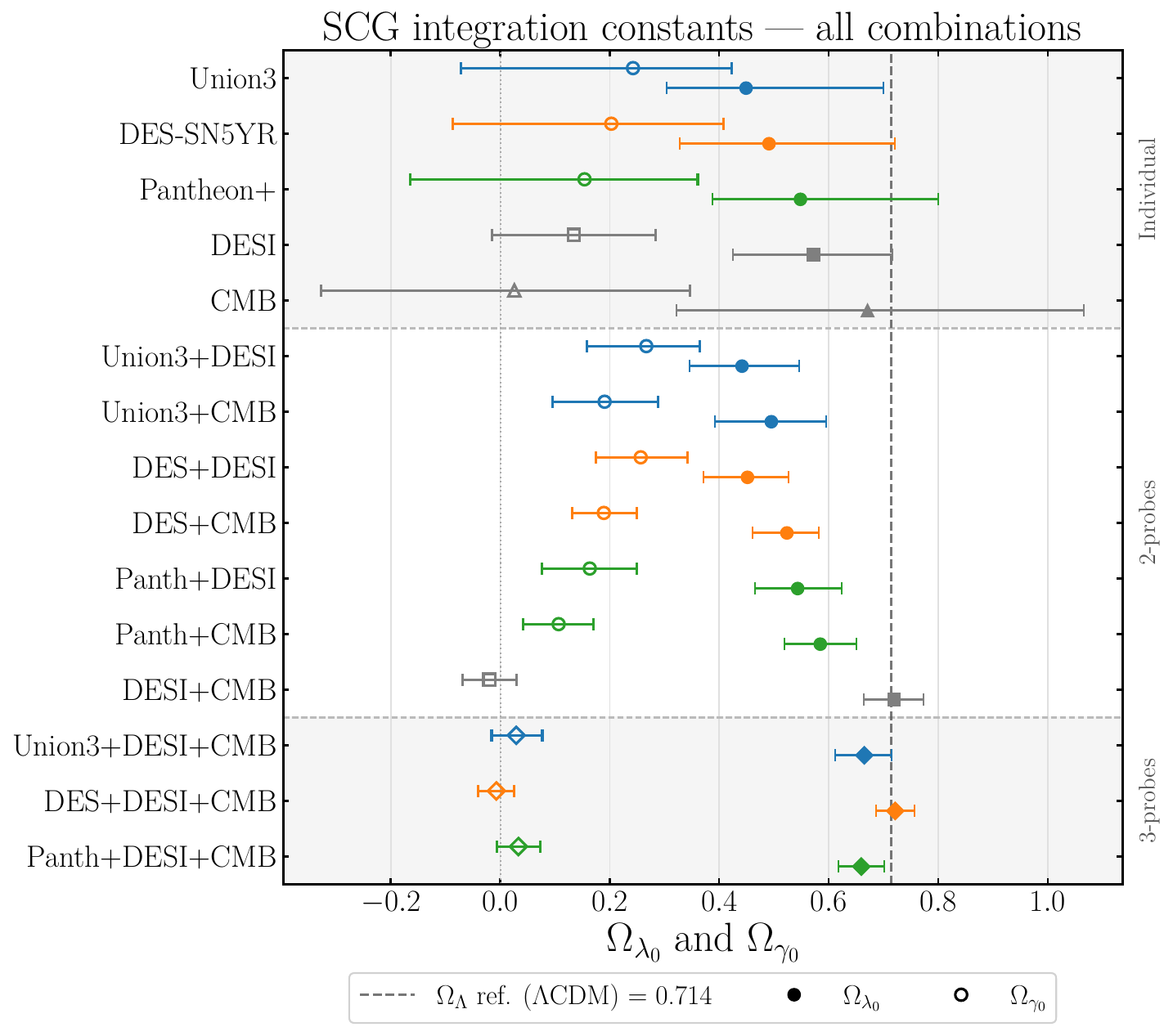}
    \caption{Marginalized constraints on the two SCG integration constants, $\Omega_{\lambda0}$ and $\Omega_{\gamma0}$. The parameter $\Omega_{\lambda0}$ represents the cosmological-constant-like contribution corresponding to the $\Lambda$CDM sector, while $\Omega_{\gamma0}$ represents the additional SCG contribution. The markers and horizontal error bars show the posterior medians and 68\% credible intervals, respectively. The vertical dashed line marks the reference $\Lambda$CDM value, $\Omega_{\Lambda0}=0.714$.}
    \label{fig:lambda_lcdm_vs_scg}
\end{figure}

The most notable change occurs after combining all three probes. The Union3-, DES-SN5YR-, and Pantheon+-based posteriors become substantially narrower and cluster around $\Omega_{\gamma0}=0$, showing that the joint SNe~Ia$+$DESI$+$CMB information breaks the degeneracies present in the individual- and two-probe analyses. Their 68\% credible regions overlap, and the pairwise differences between their medians are smaller than approximately $0.8\sigma$ when estimated using the quadrature sum of their symmetrized marginal uncertainties. These ratios are only descriptive because the analyses share DESI and CMB information and the supernova samples may also be correlated. Nevertheless, the strong marginal overlap indicates that the inferred three-probe constraint on $\Omega_{\gamma0}$ is only weakly sensitive to the choice of SNe~Ia compilation. Figure~\ref{fig:tension_3probes} therefore emphasizes the progressive contraction and reshaping of the posterior, while Fig.~\ref{fig:lambda_lcdm_vs_scg} provides the corresponding numerical credible intervals and their relation to the cosmological-constant-like contribution $\Omega_{\lambda0}$.

Figure~\ref{fig:lambda_lcdm_vs_scg} compares the cosmological-constant-like contribution $\Omega_{\lambda0}$ with the genuinely SCG contribution $\Omega_{\gamma0}$ for all 15 data combinations. For the five individual probes, the posterior on $\Omega_{\gamma0}$ remains broad, with $1\sigma$ widths ranging from approximately $0.13$ to $0.32$, and none of these configurations separates $\Omega_{\gamma0}$ from zero by more than $1\sigma$. The corresponding constraints on $\Omega_{\lambda0}$ are also broad, reflecting substantial degeneracies among the background parameters. Combining each SNe~Ia compilation with either DESI BAO or compressed CMB information consistently shifts the posterior toward positive $\Omega_{\gamma0}$ and, correspondingly, toward values of $\Omega_{\lambda0}$ below the $\Lambda$CDM reference $\Omega_{\Lambda0}=0.714$. Specifically, the two-probe combinations give $\Omega_{\gamma0}=0.267^{+0.097}_{-0.109}$ for Union3+DESI, $0.191^{+0.097}_{-0.096}$ for Union3+CMB, $0.257^{+0.086}_{-0.082}$ for DES+DESI, $0.189^{+0.061}_{-0.057}$ for DES+CMB, $0.164^{+0.086}_{-0.087}$ for Pantheon+DESI, and $0.107^{+0.063}_{-0.065}$ for Pantheon+CMB. These shifts correspond to approximate Gaussian departures from zero between $1.7\sigma$ and $3.2\sigma$, with the largest values obtained for the DES-SN5YR combinations, although they should not be interpreted as calibrated detection significances. In contrast, the SNe-independent DESI+CMB combination yields $\Omega_{\gamma0}=-0.020\pm0.049$, fully consistent with zero, and places $\Omega_{\lambda0}$ close to the $\Lambda$CDM value. The addition of all three probes removes the positive shift and further reduces the uncertainties, giving $\Omega_{\gamma0}=0.029^{+0.048}_{-0.045}$ for Union3+DESI+CMB, $-0.007^{+0.033}_{-0.034}$ for DES+DESI+CMB, and $0.033\pm0.039$ for Pantheon+DESI+CMB. All three results are consistent with the $\Lambda$CDM limit $\Omega_{\gamma0}=0$ within $1\sigma$, while their corresponding $\Omega_{\lambda0}$ constraints move toward the reference value. Overall, the figure shows that the apparent preference for a positive SCG contribution arises mainly in the SNe~Ia plus one complementary-probe combinations, whereas the complete SNe~Ia+BAO+CMB data combinations recover a predominantly cosmological-constant-like contribution with no statistically significant evidence for $\Omega_{\gamma0}\neq0$.

\begin{table*}[!t]
\centering
\caption{Bayes factors of SCG versus $\Lambda$CDM obtained with \texttt{UltraNest}. Quoted uncertainties on $\Delta\ln\mathcal{Z}$ are propagated quadratically from the individual $\ln\mathcal{Z}$ uncertainties. Positive $\Delta\ln\mathcal{Z}$ favors SCG; negative values favor $\Lambda$CDM.}
\label{tab:bayes_factors}
\begin{ruledtabular}
\begin{tabular}{lrrrll}
  Combination & $\ln\mathcal{Z}_{\mathrm{SCG}}$ & $\ln\mathcal{Z}_{\Lambda\mathrm{CDM}}$ & $\Delta\ln\mathcal{Z}$ & Strength & Favors \\
\hline
  Union3                  & $-15.11\pm0.16$  & $-15.11\pm0.22$  & $+0.00\pm0.27$ & inconclusive & --- \\
  DES-SN5YR               & $-826.77\pm0.19$ & $-827.23\pm0.47$ & $+0.46\pm0.51$ & inconclusive & SCG \\
  Pantheon$+$             & $-708.66\pm0.33$ & $-709.64\pm0.27$ & $+0.98\pm0.43$ & inconclusive & SCG \\
  DESI                    & $-12.19\pm0.18$  & $-11.49\pm0.19$  & $-0.71\pm0.26$ & inconclusive & $\Lambda$CDM \\
  CMB               & $-13.52\pm0.21$  & $-13.26\pm0.80$  & $-0.26\pm0.83$ & inconclusive & $\Lambda$CDM \\
\hline
  Union3$+$DESI           & $-25.45\pm0.29$  & $-27.57\pm0.44$  & $+2.11\pm0.53$ & moderate     & SCG \\
  Union3$+$CMB            & $-28.39\pm0.32$  & $-28.93\pm0.56$  & $+0.54\pm0.64$ & inconclusive & SCG \\
  DES-SN5YR$+$DESI        & $-836.79\pm0.23$ & $-840.68\pm0.49$ & $+3.88\pm0.54$ & strong       & SCG \\
  DES-SN5YR$+$CMB         & $-843.06\pm0.16$ & $-846.51\pm0.81$ & $+3.45\pm0.82$ & strong       & SCG \\
  Pantheon$+$$+$DESI      & $-719.00\pm0.31$ & $-720.86\pm1.02$ & $+1.86\pm1.07$ & moderate     & SCG \\
  Pantheon$+$$+$CMB       & $-721.36\pm0.35$ & $-721.95\pm0.93$ & $+0.59\pm1.00$ & inconclusive & SCG \\
  DESI$+$CMB              & $-21.59\pm0.32$  & $-19.75\pm0.66$  & $-1.84\pm0.74$ & moderate     & $\Lambda$CDM \\
\hline
  Union3$+$DESI$+$CMB       & $-37.27\pm0.48$  & $-35.37\pm0.64$  & $-1.90\pm0.80$ & moderate & $\Lambda$CDM \\
  DES-SN5YR$+$DESI$+$CMB    & $-862.73\pm0.41$ & $-860.18\pm0.75$ & $-2.54\pm0.86$ & strong   & $\Lambda$CDM \\
  Pantheon$+$$+$DESI$+$CMB  & $-729.42\pm0.31$ & $-728.29\pm0.99$ & $-1.13\pm1.04$ & moderate & $\Lambda$CDM \\
\end{tabular}
\end{ruledtabular}
\end{table*}

\subsection{Bayesian model comparison}
\label{sec:results:bayes}

To complement the parameter-level analysis we compare SCG against $\Lambda$CDM at the model level via the Bayes factor. We adopt the Jeffreys scale~\cite{Jeffreys:1961,Trotta:2008qt, Trotta:2017wnx}: $|\ln B| < 1$ is regarded as inconclusive, $1 \leq |\ln B| < 2.5$ as moderate, $2.5 \leq |\ln B| < 5$ as strong, and $|\ln B| \geq 5$ as decisive evidence in favor of the model with the larger evidence. Results for all 15 dataset combinations are collected in Table~\ref{tab:bayes_factors}.

Two complementary patterns emerge from Table~\ref{tab:bayes_factors}. First, the sign of $\Delta\ln\mathcal{Z}$ is not the same in every configuration: SCG is the preferred model in 10 of the 15 combinations, while $\Lambda$CDM is preferred in the remaining~5. SCG is favored by all five SNe~Ia-only configurations and by every two-probe combination that contains an SNe~Ia compilation, with the strongest pulls coming from DES-SN5YR$+$DESI ($\Delta\ln\mathcal{Z} = +3.88$) and DES-SN5YR$+$CMB ($+3.45$)---both reaching the ``strong'' level on the Jeffreys scale (Table~\ref{tab:bayes_factors}). Conversely, $\Lambda$CDM is favored by DESI alone, by CMB alone, by the DESI$+$CMB pair (which contains no SNe~Ia), and, most importantly, by all three full SN$+$DESI$+$CMB combinations, with the largest support against SCG provided by DES-SN5YR$+$DESI$+$CMB ($\Delta\ln\mathcal{Z} = -2.54$, strong).

The change of sign as the third probe is added mirrors the parameter-level finding of Sec.~\ref{sec:results:3probe}: in two-probe configurations SCG can absorb part of the residuals into a non-zero $\Omega_{\gamma_0}$, but in three-probe configurations the data are individually consistent with $\Omega_{\gamma_0} = 0$ and the additional flexibility offered by SCG no longer pays its way.

A caveat on the interpretation of these numbers is in order. Both the BAO observables released by DESI~DR2 and the compressed CMB likelihood we use are derived under assumptions that lean on the $\Lambda$CDM template, namely the fitting formula adopted for the sound horizon at the drag epoch $r_d$ in the BAO pipeline, and the specific choice of geometric observables $[\theta_*,\,\omega_b,\,\omega_m]$ used to summarize the CMB. A full Boltzmann-code analysis of the primary CMB anisotropy spectrum, together with a BAO reconstruction performed self consistently in SCG, could in principle shift the evidences reported here. Such an analysis lies beyond the scope of the present work, but our results should be read with this potential $\Lambda$CDM-leaning bias in mind: the disfavoring of SCG by the three-probe combinations represents a conservative bound on the model.

\section{Conclusions}\label{sec:conclusions}

We have tested the homogeneous cosmological predictions of Schouten--Codazzi gravity using 15 combinations of SNe~Ia, DESI BAO, and compressed CMB information. At the background level, SCG contains an additional geometrical contribution with the fixed effective equation of state $w_{\gamma_0}=-2/3$, while the $\Lambda$CDM expansion history is recovered in the limit $\Omega_{\gamma_0}=0$. The inferred contribution depends appreciably on the combination of probes. Each SNe~Ia compilation combined with either DESI or CMB information shifts the marginalized posterior toward positive $\Omega_{\gamma_0}$, producing approximate Gaussian departures from zero between $1.7\sigma$ and $3.2\sigma$. These values describe one-dimensional marginal posteriors and should not be interpreted as calibrated detection significances. Once SNe~Ia, DESI, and CMB information are combined simultaneously, the uncertainty decreases and all three constraints become consistent with $\Omega_{\gamma_0}=0$ within their 68\% credible intervals.

The Bayesian model comparison exhibits the same dependence on the choice and number of probes. SCG has the larger central evidence in 10 of the 15 combinations, including every two-probe combination formed from an SNe~Ia compilation and one complementary probe. The strongest support occurs for DES-SN5YR$+$DESI and DES-SN5YR$+$CMB, for which $\Delta\ln\mathcal{Z}=+3.88$ and $+3.45$, respectively, corresponding to strong evidence on the adopted Jeffreys scale. By contrast, DESI alone, CMB alone, the SNe-independent DESI$+$CMB combination, and all three SNe~Ia$+$DESI$+$CMB combinations favor $\Lambda$CDM. For the three-probe combinations, the central values span $-2.54\leq\Delta\ln\mathcal{Z}\leq-1.13$, ranging from moderate to strong support for $\Lambda$CDM according to the same scale.

The change in both the parameter constraints and the Bayesian evidence suggests that the positive SCG contribution preferred by some two-probe combinations is reduced when the remaining information is included. In particular, the addition of the mainly early-time CMB information, together with BAO constraints on the distance scale, helps break the background-parameter degeneracies that permit a larger $\Omega_{\gamma_0}$. The resulting joint posteriors move toward the $\Lambda$CDM limit, where the additional freedom of SCG is no longer sufficiently supported to compensate for its larger parameter volume. This interpretation should nevertheless remain cautious because the compressed CMB observables and parts of the BAO calibration were constructed using assumptions closely connected to a $\Lambda$CDM background. Such compression may mildly favor the region near $\Omega_{\gamma_0}=0$, although the magnitude and even the direction of the resulting shift cannot be established without a fully self-consistent analysis.

The inferred values of $H_0$ also remain close to those obtained in the corresponding $\Lambda$CDM fits. Some combinations, particularly those involving DES-SN5YR, peak between the Planck and SH0ES determinations and therefore marginally reduce the discrepancy. Nevertheless, SCG does not provide a substantial or dataset-independent resolution of the $H_0$ tension within the present background-level analysis.

These conclusions are conditional on the likelihood approximations and adopted priors. A compressed CMB likelihood extracted within $\Lambda$CDM must be validated before it can be regarded as a lossless representation of the information relevant to SCG. The BAO extraction and sound-horizon calibration introduce additional assumptions whose applicability should likewise be examined. Consequently, the present Bayes factors should be interpreted as evidence within the adopted effective likelihoods rather than as a definitive model comparison based on the complete cosmological data.

A more conclusive test requires a consistent perturbative formulation of SCG. The background scaling $w_{\gamma_0}=-2/3$ alone does not determine the evolution of density perturbations, gravitational potentials, structure growth, weak lensing, or the integrated Sachs--Wolfe effect. Deriving these predictions and implementing them in a Boltzmann solver would permit a self-consistent analysis of the full CMB anisotropy spectra, BAO observables, and large-scale-structure data. Such an analysis would determine whether the geometrical SCG contribution can be observationally distinguished from $\Lambda$CDM or from a phenomenological dark-energy component with the same homogeneous expansion history.

\section*{Data and code availability}

Numerical data, configuration files, selected validation scripts, and derived products supporting this work are publicly available at \url{https://github.com/shan1525/Cotton_gravity}. No proprietary observational data were used.

\begin{acknowledgments}
AF and AV acknowledge the support from all of the members of FISMAT, Universidad Iberoamericana. SN acknowledges support form Postdoctoral grant from Secretar\'ia de Ciencia, Humanidades, Tecnolog\'ia e Innovaci\'on SECIHTI. SAA acknowledges the support of the DGAPA postdoctoral fellowship program at ICF-UNAM, Mexico. JAV acknowledges support from FOSEC SEP-CONACYT Ciencia B\'asica A1-S-21925, UNAM-DGAPA-PAPIIT IN109126, IN110325 and C\'atedra de Investigaci\'on Marcos Moshinsky. JCH acknowledges support from UNAM-PAPIIT grant IN114626 "Estudios de ANPs y Contribuciones In-Kind a la colaboración LSST-México".

\end{acknowledgments}


\bibliography{bibfile}

@article{Perivolaropoulos:2021jda,
 archiveprefix = {arXiv},
 author = {Perivolaropoulos, Leandros and Skara, Foteini},
 doi = {10.1016/j.newar.2022.101659},
 eprint = {2105.05208},
 journal = {New Astron. Rev.},
 pages = {101659},
 primaryclass = {astro-ph.CO},
 title = {{Challenges for \ensuremath{\Lambda}CDM: An update}},
 volume = {95},
 year = {2022}
}

@article{Abdalla:2022yfr,
 archiveprefix = {arXiv},
 author = {Abdalla, Elcio and others},
 doi = {10.1016/j.jheap.2022.04.002},
 eprint = {2203.06142},
 journal = {JHEAp},
 pages = {49--211},
 primaryclass = {astro-ph.CO},
 reportnumber = {FERMILAB-CONF-22-192-SCD},
 title = {{Cosmology intertwined: A review of the particle physics, astrophysics, and cosmology associated with the cosmological tensions and anomalies}},
 volume = {34},
 year = {2022}
}

@article{DiValentino:2022fjm,
 author = {Di Valentino, Eleonora},
 doi = {10.3390/universe8080399},
 journal = {Universe},
 number = {8},
 pages = {399},
 title = {{Challenges of the Standard Cosmological Model}},
 volume = {8},
 year = {2022}
}

@inproceedings{Peebles:2024txt,
    author = "Peebles, P. J. E.",
    title = "{Status of the LambdaCDM theory: supporting evidence and anomalies}",
    eprint = "2405.18307",
    archivePrefix = "arXiv",
    primaryClass = "astro-ph.CO",
    month = "5",
    year = "2024"
}

@misc{DiValentino:2025sru,
    author = "Di Valentino, Eleonora and others",
    title = "{The CosmoVerse White Paper: Addressing observational tensions in cosmology with systematics and fundamental physics}",
    eprint = "2504.01669",
    archivePrefix = "arXiv",
    primaryClass = "astro-ph.CO",
    month = "4",
    year = "2025"
}

@article{DiValentino:2020vhf,
    author = "Di Valentino, Eleonora and others",
    title = "{Snowmass2021 - Letter of interest cosmology intertwined I: Perspectives for the next decade}",
    eprint = "2008.11283",
    archivePrefix = "arXiv",
    primaryClass = "astro-ph.CO",
    reportNumber = "FERMILAB-PUB-21-591-PPD",
    doi = "10.1016/j.astropartphys.2021.102606",
    journal = "Astropart. Phys.",
    volume = "131",
    pages = "102606",
    year = "2021"
}

@article{DiValentino:2020vvd,
    author = "Di Valentino, Eleonora and others",
    title = "{Cosmology Intertwined III: $f \sigma_8$ and $S_8$}",
    eprint = "2008.11285",
    archivePrefix = "arXiv",
    primaryClass = "astro-ph.CO",
    reportNumber = "FERMILAB-PUB-20-495-AE",
    doi = "10.1016/j.astropartphys.2021.102604",
    journal = "Astropart. Phys.",
    volume = "131",
    pages = "102604",
    year = "2021"
}

@article{DiValentino:2020srs,
    author = "Di Valentino, Eleonora and others",
    title = "{Snowmass2021 - Letter of interest cosmology intertwined IV: The age of the universe and its curvature}",
    eprint = "2008.11286",
    archivePrefix = "arXiv",
    primaryClass = "astro-ph.CO",
    reportNumber = "FERMILAB-PUB-21-592-PPD",
    doi = "10.1016/j.astropartphys.2021.102607",
    journal = "Astropart. Phys.",
    volume = "131",
    pages = "102607",
    year = "2021"
}

@article{DiValentino:2020zio,
    author = "Di Valentino, Eleonora and others",
    title = "{Snowmass2021 - Letter of interest cosmology intertwined II: The hubble constant tension}",
    eprint = "2008.11284",
    archivePrefix = "arXiv",
    primaryClass = "astro-ph.CO",
    reportNumber = "FERMILAB-PUB-21-590-PPD",
    doi = "10.1016/j.astropartphys.2021.102605",
    journal = "Astropart. Phys.",
    volume = "131",
    pages = "102605",
    year = "2021"
}

@article{Akarsu:2024qiq,
 archiveprefix = {arXiv},
 author = {Akarsu, \"Ozg\"ur and Colg\'ain, Eoin \'O. and Sen, Anjan A. and Sheikh-Jabbari, M. M.},
 doi = {10.3390/universe10080305},
 eprint = {2402.04767},
 journal = {Universe},
 number = {8},
 pages = {305},
 primaryclass = {astro-ph.CO},
 title = {{\ensuremath{\Lambda}CDM Tensions: Localising Missing Physics through Consistency Checks}},
 volume = {10},
 year = {2024}
}

@article{Verde:2019ivm,
 archiveprefix = {arXiv},
 author = {Verde, L. and Treu, T. and Riess, A. G.},
 doi = {10.1038/s41550-019-0902-0},
 eprint = {1907.10625},
 journal = {Nature Astron.},
 pages = {891},
 primaryclass = {astro-ph.CO},
 title = {{Tensions between the Early and the Late Universe}},
 volume = {3},
 year = {2019}
}

@article{DiValentino:2021izs,
 archiveprefix = {arXiv},
 author = {Di Valentino, Eleonora and Mena, Olga and Pan, Supriya and Visinelli, Luca and Yang, Weiqiang and Melchiorri, Alessandro and Mota, David F. and Riess, Adam G. and Silk, Joseph},
 doi = {10.1088/1361-6382/ac086d},
 eprint = {2103.01183},
 journal = {Class. Quant. Grav.},
 number = {15},
 pages = {153001},
 primaryclass = {astro-ph.CO},
 reportnumber = {IPPP/20/108},
 title = {{In the realm of the Hubble tension\textemdash{}a review of solutions}},
 volume = {38},
 year = {2021}
}

@article{Shah:2021onj,
    author = "Shah, Paul and Lemos, Pablo and Lahav, Ofer",
    title = "{A buyer\textquoteright{}s guide to the Hubble constant}",
    eprint = "2109.01161",
    archivePrefix = "arXiv",
    primaryClass = "astro-ph.CO",
    doi = "10.1007/s00159-021-00137-4",
    journal = "Astron. Astrophys. Rev.",
    volume = "29",
    number = "1",
    pages = "9",
    year = "2021"
}

@eprint{Kamionkowski:2022pkx,
    author = "Kamionkowski, Marc and Riess, Adam G.",
    title = "{The Hubble Tension and Early Dark Energy}",
    journal = "",
    eprint = "2211.04492",
    archivePrefix = "arXiv",
    primaryClass = "astro-ph.CO",
    month = "11",
    year = "2022"
}

@misc{Verde:2023lmm,
 archiveprefix = {arXiv},
 author = {Verde, Licia and Sch\"oneberg, Nils and Gil-Mar\'\i{}n, H\'ector},
 eprint = {2311.13305},
 month = {11},
 primaryclass = {astro-ph.CO},
 title = {{A tale of many $H_0$}},
 year = {2023}
}

@misc{Perivolaropoulos:2024yxv,
 archiveprefix = {arXiv},
 author = {Perivolaropoulos, Leandros},
 eprint = {2408.11031},
 month = {8},
 primaryclass = {astro-ph.CO},
 title = {{Hubble Tension or Distance Ladder Crisis?}},
 year = {2024}
}

@article{Planck:2018vyg,
    author = "Aghanim, N. and others",
    collaboration = "Planck",
    title = "{Planck 2018 results. VI. Cosmological parameters}",
    eprint = "1807.06209",
    archivePrefix = "arXiv",
    primaryClass = "astro-ph.CO",
    doi = "10.1051/0004-6361/201833910",
    journal = "Astron. Astrophys.",
    volume = "641",
    pages = "A6",
    year = "2020",
    note = "[Erratum: Astron.Astrophys. 652, C4 (2021)]"
}

@article{SPT-3G:2022hvq,
    author = "Balkenhol, L. and others",
    collaboration = "SPT-3G",
    title = "{Measurement of the CMB temperature power spectrum and constraints on cosmology from the SPT-3G 2018 TT, TE, and EE dataset}",
    eprint = "2212.05642",
    archivePrefix = "arXiv",
    primaryClass = "astro-ph.CO",
    reportNumber = "FERMILAB-PUB-22-953-PPD",
    doi = "10.1103/PhysRevD.108.023510",
    journal = "Phys. Rev. D",
    volume = "108",
    number = "2",
    pages = "023510",
    year = "2023"
}

@misc{ACT:2025fju,
    author = "Louis, Thibaut and others",
    collaboration = "ACT",
    title = "{The Atacama Cosmology Telescope: DR6 Power Spectra, Likelihoods and $\Lambda$CDM Parameters}",
    eprint = "2503.14452",
    archivePrefix = "arXiv",
    primaryClass = "astro-ph.CO",
    reportNumber = "FERMILAB-PUB-25-0071-PPD",
    month = "3",
    year = "2025"
}

@article{Freedman:2020dne,
 archiveprefix = {arXiv},
 author = {Freedman, Wendy L. and Madore, Barry F. and Hoyt, Taylor and Jang, In Sung and Beaton, Rachael and Lee, Myung Gyoon and Monson, Andrew and Neeley, Jill and Rich, Jeffrey},
 doi = {10.3847/1538-4357/ab7339},
 eprint = {2002.01550},
 journal = {Astrophys. J.},
 month = {2},
 number = {1},
 pages = {57},
 primaryclass = {astro-ph.GA},
 title = {{Calibration of the Tip of the Red Giant Branch (TRGB)}},
 volume = {891},
 year = {2020}
}

@article{Birrer:2020tax,
 archiveprefix = {arXiv},
 author = {Birrer, S. and others},
 doi = {10.1051/0004-6361/202038861},
 eprint = {2007.02941},
 journal = {Astron. Astrophys.},
 pages = {A165},
 primaryclass = {astro-ph.CO},
 title = {{TDCOSMO - IV. Hierarchical time-delay cosmography \textendash{} joint inference of the Hubble constant and galaxy density profiles}},
 volume = {643},
 year = {2020}
}

@article{Wu:2021jyk,
 archiveprefix = {arXiv},
 author = {Wu, Qin and Zhang, Guo-Qiang and Wang, Fa-Yin},
 doi = {10.1093/mnrasl/slac022},
 eprint = {2108.00581},
 journal = {Mon. Not. Roy. Astron. Soc.},
 note = {[Erratum: Mon.Not.Roy.Astron.Soc. 531, L8 (2024)]},
 number = {1},
 pages = {L1--L5},
 primaryclass = {astro-ph.CO},
 title = {{An 8~per\,cent determination of the Hubble constant from localized fast radio bursts}},
 volume = {515},
 year = {2022}
}

@eprint{Anderson:2023aga,
    author = "Anderson, Richard I. and Koblischke, Nolan W. and Eyer, Laurent",
    title = "{Reconciling astronomical distance scales with variable red giant stars}",
    journal = "",
    eprint = "2303.04790",
    archivePrefix = "arXiv",
    primaryClass = "astro-ph.CO",
    month = "3",
    year = "2023"
}

@eprint{Scolnic:2023mrv,
    author = "Scolnic, D. and Riess, A. G. and Wu, J. and Li, S. and Anand, G. S. and Beaton, R. and Casertano, S. and Anderson, R. and Dhawan, S. and Ke, X.",
    title = "{CATS: The Hubble Constant from Standardized TRGB and Type Ia Supernova Measurements}",
    journal = "",
    eprint = "2304.06693",
    archivePrefix = "arXiv",
    primaryClass = "astro-ph.CO",
    month = "4",
    year = "2023"
}

@article{Jones:2022mvo,
 archiveprefix = {arXiv},
 author = {Jones, D. O. and others},
 doi = {10.3847/1538-4357/ac755b},
 eprint = {2201.07801},
 journal = {Astrophys. J.},
 number = {2},
 pages = {172},
 primaryclass = {astro-ph.CO},
 title = {{Cosmological Results from the RAISIN Survey: Using Type Ia Supernovae in the Near Infrared as a Novel Path to Measure the Dark Energy Equation of State}},
 volume = {933},
 year = {2022}
}

@article{Anand:2021sum,
 archiveprefix = {arXiv},
 author = {Anand, Gagandeep S. and Tully, R. Brent and Rizzi, Luca and Riess, Adam G. and Yuan, Wenlong},
 doi = {10.3847/1538-4357/ac68df},
 eprint = {2108.00007},
 journal = {Astrophys. J.},
 number = {1},
 pages = {15},
 primaryclass = {astro-ph.CO},
 title = {{Comparing Tip of the Red Giant Branch Distance Scales: An Independent Reduction of the Carnegie-Chicago Hubble Program and the Value of the Hubble Constant}},
 volume = {932},
 year = {2022}
}

@article{Freedman:2021ahq,
    author = "Freedman, Wendy L.",
    title = "{Measurements of the Hubble Constant: Tensions in Perspective}",
    eprint = "2106.15656",
    archivePrefix = "arXiv",
    primaryClass = "astro-ph.CO",
    doi = "10.3847/1538-4357/ac0e95",
    journal = "Astrophys. J.",
    volume = "919",
    number = "1",
    pages = "16",
    year = "2021"
}

@article{Uddin:2023iob,
 archiveprefix = {arXiv},
 author = {Uddin, Syed A. and others},
 doi = {10.3847/1538-4357/ad3e63},
 eprint = {2308.01875},
 journal = {Astrophys. J.},
 number = {1},
 pages = {72},
 primaryclass = {astro-ph.CO},
 title = {{Carnegie Supernova Project I and II: Measurements of H $_{0}$ Using Cepheid, Tip of the Red Giant Branch, and Surface Brightness Fluctuation Distance Calibration to Type Ia Supernovae*}},
 volume = {970},
 year = {2024}
}

@article{Huang:2023frr,
 archiveprefix = {arXiv},
 author = {Huang, Caroline D. and others},
 doi = {10.3847/1538-4357/ad1ff8},
 eprint = {2312.08423},
 journal = {Astrophys. J.},
 number = {2},
 pages = {83},
 primaryclass = {astro-ph.CO},
 title = {{The Mira Distance to M101 and a 4\% Measurement of H $_{0}$}},
 volume = {963},
 year = {2024}
}

@article{Pesce:2020xfe,
    author = "Pesce, D. W. and others",
    title = "{The Megamaser Cosmology Project. XIII. Combined Hubble constant constraints}",
    eprint = "2001.09213",
    archivePrefix = "arXiv",
    primaryClass = "astro-ph.CO",
    doi = "10.3847/2041-8213/ab75f0",
    journal = "Astrophys. J. Lett.",
    volume = "891",
    number = "1",
    pages = "L1",
    year = "2020"
}

@article{Kourkchi:2020iyz,
 archiveprefix = {arXiv},
 author = {Kourkchi, Ehsan and Tully, R. Brent and Anand, Gagandeep S. and Courtois, Helene M. and Dupuy, Alexandra and Neill, James D. and Rizzi, Luca and Seibert, Mark},
 doi = {10.3847/1538-4357/ab901c},
 eprint = {2004.14499},
 journal = {Astrophys. J.},
 number = {1},
 pages = {3},
 primaryclass = {astro-ph.GA},
 title = {{Cosmicflows-4: The Calibration of Optical and Infrared Tully\textendash{}Fisher Relations}},
 volume = {896},
 year = {2020}
}

@article{Schombert:2020pxm,
 archiveprefix = {arXiv},
 author = {Schombert, James and McGaugh, Stacy and Lelli, Federico},
 doi = {10.3847/1538-3881/ab9d88},
 eprint = {2006.08615},
 journal = {Astron. J.},
 number = {2},
 pages = {71},
 primaryclass = {astro-ph.CO},
 title = {{Using the Baryonic Tully\textendash{}Fisher Relation to Measure H o}},
 volume = {160},
 year = {2020}
}

@article{Blakeslee:2021rqi,
    author = "Blakeslee, John P. and Jensen, Joseph B. and Ma, Chung-Pei and Milne, Peter A. and Greene, Jenny E.",
    title = "{The Hubble Constant from Infrared Surface Brightness Fluctuation Distances}",
    eprint = "2101.02221",
    archivePrefix = "arXiv",
    primaryClass = "astro-ph.CO",
    doi = "10.3847/1538-4357/abe86a",
    journal = "Astrophys. J.",
    volume = "911",
    number = "1",
    pages = "65",
    year = "2021"
}

@article{deJaeger:2022lit,
    author = "de Jaeger, T. and Galbany, L. and Riess, A. G. and Stahl, B. E. and Shappee, B. J. and Filippenko, A. V. and Zheng, W.",
    title = "{A 5~per\,cent measurement of the Hubble\textendash{}Lema\^\i{}tre constant from Type II supernovae}",
    eprint = "2203.08974",
    archivePrefix = "arXiv",
    primaryClass = "astro-ph.CO",
    doi = "10.1093/mnras/stac1661",
    journal = "Mon. Not. Roy. Astron. Soc.",
    volume = "514",
    number = "3",
    pages = "4620--4628",
    year = "2022"
}

@misc{Murakami:2023xuy,
    author = "Murakami, Yukei S. and Riess, Adam G. and Stahl, Benjamin E. and Kenworthy, W. D'Arcy and Pluck, Dahne-More A. and Macoretta, Antonella and Brout, Dillon and Jones, David O. and Scolnic, Dan M. and Filippenko, Alexei V.",
    title = "{Leveraging SN Ia spectroscopic similarity to improve the measurement of $H_0$}",
    journal = "",
    eprint = "2306.00070",
    archivePrefix = "arXiv",
    primaryClass = "astro-ph.CO",
    month = "5",
    year = "2023"
}

@article{Breuval:2024lsv,
 archiveprefix = {arXiv},
 author = {Breuval, Louise and Riess, Adam G. and Casertano, Stefano and Yuan, Wenlong and Macri, Lucas M. and Romaniello, Martino and Murakami, Yukei S. and Scolnic, Daniel and Anand, Gagandeep S. and Soszy\'nski, Igor},
 doi = {10.3847/1538-4357/ad630e},
 eprint = {2404.08038},
 journal = {Astrophys. J.},
 number = {1},
 pages = {30},
 primaryclass = {astro-ph.CO},
 title = {{Small Magellanic Cloud Cepheids Observed with the Hubble Space Telescope Provide a New Anchor for the SH0ES Distance Ladder}},
 volume = {973},
 year = {2024}
}

@misc{Freedman:2024eph,
 archiveprefix = {arXiv},
 author = {Freedman, Wendy L. and Madore, Barry F. and Jang, In Sung and Hoyt, Taylor J. and Lee, Abigail J. and Owens, Kayla A.},
 eprint = {2408.06153},
 month = {8},
 primaryclass = {astro-ph.CO},
 title = {{Status Report on the Chicago-Carnegie Hubble Program (CCHP): Three Independent Astrophysical Determinations of the Hubble Constant Using the James Webb Space Telescope}},
 year = {2024}
}

@misc{Riess:2024vfa,
 archiveprefix = {arXiv},
 author = {Riess, A. G. and others},
 eprint = {2408.11770},
 month = {8},
 primaryclass = {astro-ph.CO},
 title = {{JWST Validates HST Distance Measurements: Selection of Supernova Subsample Explains Differences in JWST Estimates of Local H0}},
 year = {2024}
}

@misc{Vogl:2024bum,
 archiveprefix = {arXiv},
 author = {Vogl, Christian and others},
 eprint = {2411.04968},
 month = {11},
 primaryclass = {astro-ph.CO},
 title = {{No rungs attached: A distance-ladder free determination of the Hubble constant through type II supernova spectral modelling}},
 year = {2024}
}

@misc{Gao:2024kkx,
 archiveprefix = {arXiv},
 author = {Gao, D. H. and Wu, Q. and Hu, J. P. and Yi, S. X. and Zhou, X. and Wang, F. Y.},
 eprint = {2410.03994},
 month = {10},
 primaryclass = {astro-ph.CO},
 title = {{Measuring Hubble constant using localized and unlocalized fast radio bursts}},
 year = {2024}
}

@misc{Scolnic:2024hbh,
 archiveprefix = {arXiv},
 author = {Scolnic, Daniel and others},
 eprint = {2409.14546},
 month = {9},
 primaryclass = {astro-ph.CO},
 title = {{The Hubble Tension in our own Backyard: DESI and the Nearness of the Coma Cluster}},
 year = {2024}
}

@misc{Said:2024pwm,
 archiveprefix = {arXiv},
 author = {Said, Khaled and others},
 eprint = {2408.13842},
 month = {8},
 primaryclass = {astro-ph.CO},
 title = {{DESI Peculiar Velocity Survey \textendash{} Fundamental Plane}},
 year = {2024}
}

@article{Boubel:2024cqw,
 archiveprefix = {arXiv},
 author = {Boubel, Paula and Colless, Matthew and Said, Khaled and Staveley-Smith, Lister},
 doi = {10.1093/mnras/stae1925},
 eprint = {2408.03660},
 journal = {Mon. Not. Roy. Astron. Soc.},
 number = {2},
 pages = {1550--1559},
 primaryclass = {astro-ph.CO},
 title = {{An improved Tully\textendash{}Fisher estimate of H0}},
 volume = {533},
 year = {2024}
}

@misc{Scolnic:2024oth,
    author = "Scolnic, Daniel and Boubel, Paula and Byrne, Jakob and Riess, Adam G. and Anand, Gagandeep S.",
    title = "{Calibrating the Tully-Fisher Relation to Measure the Hubble Constant}",
    eprint = "2412.08449",
    archivePrefix = "arXiv",
    primaryClass = "astro-ph.CO",
    month = "12",
    year = "2024"
}

@misc{Li:2025ife,
    author = "Li, Siyang and Riess, Adam G. and Scolnic, Daniel and Casertano, Stefano and Anand, Gagandeep S.",
    title = "{JAGB 2.0: Improved Constraints on the J-region Asymptotic Giant Branch-based Hubble Constant from an Expanded Sample of JWST Observations}",
    eprint = "2502.05259",
    archivePrefix = "arXiv",
    primaryClass = "astro-ph.CO",
    month = "2",
    year = "2025"
}

@misc{Jensen:2025aai,
    author = "Jensen, Joseph B. and Blakeslee, John P. and Cantiello, Michele and Cowles, Mikaela and Anand, Gagandeep S. and Tully, R. Brent and Kourkchi, Ehsan and Raimondo, Gabriella",
    title = "{The TRGB-SBF Project. III. Refining the HST Surface Brightness Fluctuation Distance Scale Calibration with JWST}",
    eprint = "2502.15935",
    archivePrefix = "arXiv",
    primaryClass = "astro-ph.CO",
    month = "2",
    year = "2025"
}

@article{Riess:2022shoes,
  author  = {Riess, Adam G. and others},
  title   = {A Comprehensive Measurement of the Local Value of the
             {H}ubble Constant with 1\,km\,s$^{-1}$\,Mpc$^{-1}$
             Uncertainty from the {H}ubble Space Telescope and the
             {SH0ES} Team},
  journal = {Astrophys.\ J.\ Lett.},
  volume  = {934},
  pages   = {L7},
  year    = {2022},
  doi     = {10.3847/2041-8213/ac5c5b},
  eprint  = {2112.04510},
  archivePrefix = {arXiv},
  primaryClass  = {astro-ph.CO}
}

@misc{Efstathiou:2020wxn,
 archiveprefix = {arXiv},
 author = {Efstathiou, G.},
 eprint = {2007.10716},
 month = {7},
 primaryclass = {astro-ph.CO},
 title = {{A Lockdown Perspective on the Hubble Tension (with comments from the SH0ES team)}},
 year = {2020}
}

@article{Mortsell:2021nzg,
 archiveprefix = {arXiv},
 author = {Mortsell, Edvard and Goobar, Ariel and Johansson, Joel and Dhawan, Suhail},
 doi = {10.3847/1538-4357/ac756e},
 eprint = {2105.11461},
 journal = {Astrophys. J.},
 number = {2},
 pages = {212},
 primaryclass = {astro-ph.CO},
 title = {{Sensitivity of the Hubble Constant Determination to Cepheid Calibration}},
 volume = {933},
 year = {2022}
}

@article{Mortsell:2021tcx,
 archiveprefix = {arXiv},
 author = {Mortsell, Edvard and Goobar, Ariel and Johansson, Joel and Dhawan, Suhail},
 doi = {10.3847/1538-4357/ac7c19},
 eprint = {2106.09400},
 journal = {Astrophys. J.},
 number = {1},
 pages = {58},
 primaryclass = {astro-ph.CO},
 title = {{The Hubble Tension Revisited: Additional Local Distance Ladder Uncertainties}},
 volume = {935},
 year = {2022}
}

@article{Sharon:2023ioz,
 archiveprefix = {arXiv},
 author = {Sharon, Amir and Kushnir, Doron and Yuan, Wenlong and Macri, Lucas and Riess, Adam},
 doi = {10.1093/mnras/stae451},
 eprint = {2305.14435},
 journal = {Mon. Not. Roy. Astron. Soc.},
 number = {4},
 pages = {6861--6880},
 primaryclass = {astro-ph.CO},
 title = {{Reassessing the constraints from SH0ES extragalactic Cepheid amplitudes on systematic blending bias}},
 volume = {528},
 year = {2024}
}

@article{Riess:2023bfx,
 archiveprefix = {arXiv},
 author = {Riess, Adam G. and Anand, Gagandeep S. and Yuan, Wenlong and Casertano, Stefano and Dolphin, Andrew and Macri, Lucas M. and Breuval, Louise and Scolnic, Dan and Perrin, Marshall and Anderson, Richard I.},
 doi = {10.3847/2041-8213/acf769},
 eprint = {2307.15806},
 journal = {Astrophys. J. Lett.},
 number = {1},
 pages = {L18},
 primaryclass = {astro-ph.CO},
 title = {{Crowded No More: The Accuracy of the Hubble Constant Tested with High-resolution Observations of Cepheids by JWST}},
 volume = {956},
 year = {2023}
}

@article{Bhardwaj:2023mau,
 archiveprefix = {arXiv},
 author = {Bhardwaj, Anupam and others},
 doi = {10.3847/2041-8213/acf710},
 eprint = {2309.03263},
 journal = {Astrophys. J. Lett.},
 number = {1},
 pages = {L13},
 primaryclass = {astro-ph.SR},
 title = {{High-resolution Spectroscopic Metallicities of Milky Way Cepheid Standards and Their Impact on the Leavitt Law and the Hubble Constant}},
 volume = {955},
 year = {2023}
}

@misc{Brout:2023wol,
 archiveprefix = {arXiv},
 author = {Brout, Dillon and Riess, Adam},
 eprint = {2311.08253},
 month = {11},
 primaryclass = {astro-ph.CO},
 title = {{The Impact of Dust on Cepheid and Type Ia Supernova Distances}},
 year = {2023}
}

@article{Dwomoh:2023bro,
 archiveprefix = {arXiv},
 author = {Dwomoh, Arianna M. and Peterson, Erik R. and Scolnic, Daniel and Ashall, Chris and DerKacy, James M. and Do, Aaron and Johansson, Joel and Jones, David O. and Riess, Adam G. and Shappee, Benjamin J.},
 doi = {10.3847/1538-4357/ad1ff5},
 eprint = {2311.06178},
 journal = {Astrophys. J.},
 number = {1},
 pages = {90},
 primaryclass = {astro-ph.CO},
 title = {{Evaluating the Consistency of Cosmological Distances Using Supernova Siblings in the Near-infrared}},
 volume = {965},
 year = {2024}
}

@article{Riess:2024ohe,
 archiveprefix = {arXiv},
 author = {Riess, Adam G. and Anand, Gagandeep S. and Yuan, Wenlong and Casertano, Stefano and Dolphin, Andrew and Macri, Lucas M. and Breuval, Louise and Scolnic, Dan and Perrin, Marshall and Anderson, I. Richard},
 doi = {10.3847/2041-8213/ad1ddd},
 eprint = {2401.04773},
 journal = {Astrophys. J. Lett.},
 number = {1},
 pages = {L17},
 primaryclass = {astro-ph.CO},
 title = {{JWST Observations Reject Unrecognized Crowding of Cepheid Photometry as an Explanation for the Hubble Tension at 8\ensuremath{\sigma} Confidence}},
 volume = {962},
 year = {2024}
}

@article{Knox:2019rjx,
 archiveprefix = {arXiv},
 author = {Knox, Lloyd and Millea, Marius},
 doi = {10.1103/PhysRevD.101.043533},
 eprint = {1908.03663},
 journal = {Phys. Rev. D},
 number = {4},
 pages = {043533},
 primaryclass = {astro-ph.CO},
 title = {{Hubble constant hunter\textquoteright{}s guide}},
 volume = {101},
 year = {2020}
}

@article{Jedamzik:2020zmd,
    author = "Jedamzik, Karsten and Pogosian, Levon and Zhao, Gong-Bo",
    title = "{Why reducing the cosmic sound horizon alone can not fully resolve the Hubble tension}",
    eprint = "2010.04158",
    archivePrefix = "arXiv",
    primaryClass = "astro-ph.CO",
    doi = "10.1038/s42005-021-00628-x",
    journal = "Commun. in Phys.",
    volume = "4",
    pages = "123",
    year = "2021"
}

@article{Schoneberg:2021qvd,
 archiveprefix = {arXiv},
 author = {Sch\"oneberg, Nils and Franco Abell\'an, Guillermo and P\'erez S\'anchez, Andrea and Witte, Samuel J. and Poulin, Vivian and Lesgourgues, Julien},
 doi = {10.1016/j.physrep.2022.07.001},
 eprint = {2107.10291},
 journal = {Phys. Rept.},
 pages = {1--55},
 primaryclass = {astro-ph.CO},
 title = {{The H0 Olympics: A fair ranking of proposed models}},
 volume = {984},
 year = {2022}
}

@article{Khalife:2023qbu,
 archiveprefix = {arXiv},
 author = {Khalife, Ali Rida and Zanjani, Maryam Bahrami and Galli, Silvia and G\"unther, Sven and Lesgourgues, Julien and Benabed, Karim},
 doi = {10.1088/1475-7516/2024/04/059},
 eprint = {2312.09814},
 journal = {JCAP},
 pages = {059},
 primaryclass = {astro-ph.CO},
 reportnumber = {TTK-23-36},
 title = {{Review of Hubble tension solutions with new SH0ES and SPT-3G data}},
 volume = {04},
 year = {2024}
}

@article{Hu:2023jqc,
 archiveprefix = {arXiv},
 author = {Hu, Jian-Ping and Wang, Fa-Yin},
 doi = {10.3390/universe9020094},
 eprint = {2302.05709},
 journal = {Universe},
 number = {2},
 pages = {94},
 primaryclass = {astro-ph.CO},
 title = {{Hubble Tension: The Evidence of New Physics}},
 volume = {9},
 year = {2023}
}

@article{Nunes:2021ipq,
 archiveprefix = {arXiv},
 author = {Nunes, Rafael C. and Vagnozzi, Sunny},
 doi = {10.1093/mnras/stab1613},
 eprint = {2106.01208},
 journal = {Mon. Not. Roy. Astron. Soc.},
 number = {4},
 pages = {5427--5437},
 primaryclass = {astro-ph.CO},
 title = {{Arbitrating the S8 discrepancy with growth rate measurements from redshift-space distortions}},
 volume = {505},
 year = {2021}
}

@article{DiValentino:2018gcu,
 author = {Di Valentino, Eleonora and Bridle, Sarah},
 doi = {10.3390/sym10110585},
 journal = {Symmetry},
 number = {11},
 pages = {585},
 title = {{Exploring the Tension between Current Cosmic Microwave Background and Cosmic Shear Data}},
 volume = {10},
 year = {2018}
}

@article{Adil:2023jtu,
    author = {Adil, Shahnawaz A. and Akarsu, \"Ozg\"ur and Malekjani, Mohammad and Colg\'ain, Eoin \'O. and Pourojaghi, Saeed and Sen, Anjan A. and Sheikh-Jabbari, M. M.},
    title = "{S8 increases with effective redshift in \ensuremath{\Lambda}CDM cosmology}",
    eprint = "2303.06928",
    archivePrefix = "arXiv",
    primaryClass = "astro-ph.CO",
    doi = "10.1093/mnrasl/slad165",
    journal = "Mon. Not. Roy. Astron. Soc.",
    volume = "528",
    number = "1",
    pages = "L20--L26",
    year = "2023"
}

@article{Akarsu:2024hsu,
    author = {Akarsu, {\"O}zg{\"u}r and Colg{\'a}in, Eoin {\'O}. and Sen, Anjan A. and Sheikh-Jabbari, M. M.},
    title = "{Further support for $S_8$ increasing with effective redshift}",
    eprint = "2410.23134",
    archivePrefix = "arXiv",
    primaryClass = "astro-ph.CO",
    doi = "10.1093/mnrasl/slaf041",
    journal = "Mon. Not. Roy. Astron. Soc.",
    volume = "542",
    number = "1",
    pages = "L36--L42",
    year = "2025"
}

@article{DES:2021bvc,
 archiveprefix = {arXiv},
 author = {Amon, A. and others},
 collaboration = {DES},
 doi = {10.1103/PhysRevD.105.023514},
 eprint = {2105.13543},
 journal = {Phys. Rev. D},
 number = {2},
 pages = {023514},
 primaryclass = {astro-ph.CO},
 reportnumber = {FERMILAB-PUB-21-250-AE, DES-2019-0479},
 title = {{Dark Energy Survey Year 3 results: Cosmology from cosmic shear and robustness to data calibration}},
 volume = {105},
 year = {2022}
}

@article{DES:2021vln,
    author = "Secco, L. F. and others",
    collaboration = "DES",
    title = "{Dark Energy Survey Year 3 results: Cosmology from cosmic shear and robustness to modeling uncertainty}",
    eprint = "2105.13544",
    archivePrefix = "arXiv",
    primaryClass = "astro-ph.CO",
    reportNumber = "FERMILAB-PUB-21-253-AE, DES-2019-0480",
    doi = "10.1103/PhysRevD.105.023515",
    journal = "Phys. Rev. D",
    volume = "105",
    number = "2",
    pages = "023515",
    year = "2022"
}

@article{KiDS:2020suj,
    author = "Asgari, Marika and others",
    collaboration = "KiDS",
    title = "{KiDS-1000 Cosmology: Cosmic shear constraints and comparison between two point statistics}",
    eprint = "2007.15633",
    archivePrefix = "arXiv",
    primaryClass = "astro-ph.CO",
    doi = "10.1051/0004-6361/202039070",
    journal = "Astron. Astrophys.",
    volume = "645",
    pages = "A104",
    year = "2021"
}

@article{Asgari:2019fkq,
 archiveprefix = {arXiv},
 author = {Asgari, Marika and others},
 doi = {10.1051/0004-6361/201936512},
 eprint = {1910.05336},
 journal = {Astron. Astrophys.},
 pages = {A127},
 primaryclass = {astro-ph.CO},
 title = {{KiDS+VIKING-450 and DES-Y1 combined: Mitigating baryon feedback uncertainty with COSEBIs}},
 volume = {634},
 year = {2020}
}

@article{Joudaki:2019pmv,
    author = "Joudaki, S. and others",
    title = "{KiDS+VIKING-450 and DES-Y1 combined: Cosmology with cosmic shear}",
    eprint = "1906.09262",
    archivePrefix = "arXiv",
    primaryClass = "astro-ph.CO",
    doi = "10.1051/0004-6361/201936154",
    journal = "Astron. Astrophys.",
    volume = "638",
    pages = "L1",
    year = "2020"
}

@article{DAmico:2019fhj,
 archiveprefix = {arXiv},
 author = {D'Amico, Guido and Gleyzes, J\'er\^ome and Kokron, Nickolas and Markovic, Katarina and Senatore, Leonardo and Zhang, Pierre and Beutler, Florian and Gil-Mar\'\i{}n, H\'ector},
 doi = {10.1088/1475-7516/2020/05/005},
 eprint = {1909.05271},
 journal = {JCAP},
 pages = {005},
 primaryclass = {astro-ph.CO},
 title = {{The Cosmological Analysis of the SDSS/BOSS data from the Effective Field Theory of Large-Scale Structure}},
 volume = {05},
 year = {2020}
}

@article{Kilo-DegreeSurvey:2023gfr,
 archiveprefix = {arXiv},
 author = {Abbott, T. M. C. and others},
 collaboration = {Kilo-Degree Survey, DES},
 doi = {10.21105/astro.2305.17173},
 eprint = {2305.17173},
 journal = {Open J. Astrophys.},
 pages = {2305.17173},
 primaryclass = {astro-ph.CO},
 reportnumber = {FERMILAB-PUB-23-267-PPD, DES-2023-0769},
 title = {{DES Y3 + KiDS-1000: Consistent cosmology combining cosmic shear surveys}},
 volume = {6},
 year = {2023}
}

@article{Troster:2019ean,
 archiveprefix = {arXiv},
 author = {Tr\"oster, Tilman and others},
 doi = {10.1051/0004-6361/201936772},
 eprint = {1909.11006},
 journal = {Astron. Astrophys.},
 pages = {L10},
 primaryclass = {astro-ph.CO},
 title = {{Cosmology from large-scale structure: Constraining $\Lambda$CDM with BOSS}},
 volume = {633},
 year = {2020}
}

@article{Heymans:2020gsg,
 archiveprefix = {arXiv},
 author = {Heymans, Catherine and others},
 doi = {10.1051/0004-6361/202039063},
 eprint = {2007.15632},
 journal = {Astron. Astrophys.},
 pages = {A140},
 primaryclass = {astro-ph.CO},
 title = {{KiDS-1000 Cosmology: Multi-probe weak gravitational lensing and spectroscopic galaxy clustering constraints}},
 volume = {646},
 year = {2021}
}

@eprint{Dalal:2023olq,
    author = "Dalal, Roohi and others",
    title = "{Hyper Suprime-Cam Year 3 Results: Cosmology from Cosmic Shear Power Spectra}",
    journal = "",
    eprint = "2304.00701",
    archivePrefix = "arXiv",
    primaryClass = "astro-ph.CO",
    month = "4",
    year = "2023"
}

@misc{Chen:2024vvk,
 archiveprefix = {arXiv},
 author = {Chen, S. and others},
 eprint = {2407.04795},
 month = {7},
 primaryclass = {astro-ph.CO},
 reportnumber = {FERMILAB-PUB-24-0137-PPD},
 title = {{Not all lensing is low: An analysis of DESI$\times$DES using the Lagrangian Effective Theory of LSS}},
 year = {2024}
}

@misc{ACT:2024okh,
 archiveprefix = {arXiv},
 author = {Kim, Joshua and others},
 collaboration = {ACT, DESI},
 eprint = {2407.04606},
 month = {7},
 primaryclass = {astro-ph.CO},
 title = {{The Atacama Cosmology Telescope DR6 and DESI: Structure formation over cosmic time with a measurement of the cross-correlation of CMB Lensing and Luminous Red Galaxies}},
 year = {2024}
}

@misc{Sailer:2024coh,
    author = "Sailer, Noah and others",
    title = "{Cosmological constraints from the cross-correlation of DESI Luminous Red Galaxies with CMB lensing from Planck PR4 and ACT DR6}",
    eprint = "2407.04607",
    archivePrefix = "arXiv",
    primaryClass = "astro-ph.CO",
    month = "7",
    year = "2024"
}

@misc{Harnois-Deraps:2024ucb,
 archiveprefix = {arXiv},
 author = {Harnois-Deraps, Joachim and others},
 eprint = {2405.10312},
 month = {5},
 primaryclass = {astro-ph.CO},
 title = {{KiDS-1000 and DES-Y1 combined: Cosmology from peak count statistics}},
 year = {2024}
}

@article{Dvornik:2022xap,
 archiveprefix = {arXiv},
 author = {Dvornik, Andrej and others},
 doi = {10.1051/0004-6361/202245158},
 eprint = {2210.03110},
 journal = {Astron. Astrophys.},
 pages = {A189},
 primaryclass = {astro-ph.CO},
 title = {{KiDS-1000: Combined halo-model cosmology constraints from galaxy abundance, galaxy clustering and galaxy-galaxy lensing}},
 volume = {675},
 year = {2023}
}

@article{DES:2021wwk,
    author = "Abbott, T. M. C. and others",
    collaboration = "DES",
    title = "{Dark Energy Survey Year 3 results: Cosmological constraints from galaxy clustering and weak lensing}",
    eprint = "2105.13549",
    archivePrefix = "arXiv",
    primaryClass = "astro-ph.CO",
    reportNumber = "FERMILAB-PUB-21-221-AE, DES-2020-0617",
    doi = "10.1103/PhysRevD.105.023520",
    journal = "Phys. Rev. D",
    volume = "105",
    number = "2",
    pages = "023520",
    year = "2022"
}

@misc{Wright:2025xka,
    author = "Wright, Angus H. and others",
    title = "{KiDS-Legacy: Cosmological constraints from cosmic shear with the complete Kilo-Degree Survey}",
    eprint = "2503.19441",
    archivePrefix = "arXiv",
    primaryClass = "astro-ph.CO",
    month = "3",
    year = "2025"
}

@article{DESI:2024mwx,
    author = "Adame, A. G. and others",
    collaboration = "DESI",
    title = "{DESI 2024 VI: cosmological constraints from the measurements of baryon acoustic oscillations}",
    eprint = "2404.03002",
    archivePrefix = "arXiv",
    primaryClass = "astro-ph.CO",
    reportNumber = "FERMILAB-PUB-24-0154-PPD",
    doi = "10.1088/1475-7516/2025/02/021",
    journal = "JCAP",
    volume = "02",
    pages = "021",
    year = "2025"
}

@article{DESI:2024aqx,
 archiveprefix = {arXiv},
 author = {Calderon, R. and others},
 collaboration = {DESI},
 doi = {10.1088/1475-7516/2024/10/048},
 eprint = {2405.04216},
 journal = {JCAP},
 pages = {048},
 primaryclass = {astro-ph.CO},
 title = {{DESI 2024: reconstructing dark energy using crossing statistics with DESI DR1 BAO data}},
 volume = {10},
 year = {2024}
}

@misc{DESI:2024jis,
    author = "Adame, A. G. and others",
    collaboration = "DESI",
    title = "{DESI 2024 V: Full-Shape Galaxy Clustering from Galaxies and Quasars}",
    eprint = "2411.12021",
    archivePrefix = "arXiv",
    primaryClass = "astro-ph.CO",
    reportNumber = "FERMILAB-PUB-24-0847-PPD",
    month = "11",
    year = "2024"
}

@misc{DESI:2025zgx,
    author = "Abdul Karim, M. and others",
    collaboration = "DESI",
    title = "{DESI DR2 Results II: Measurements of Baryon Acoustic Oscillations and Cosmological Constraints}",
    eprint = "2503.14738",
    archivePrefix = "arXiv",
    primaryClass = "astro-ph.CO",
    reportNumber = "FERMILAB-PUB-25-0169-PPD",
    month = "3",
    year = "2025"
}

@misc{DESI:2025fii,
    author = "Lodha, K. and others",
    collaboration = "DESI",
    title = "{Extended Dark Energy analysis using DESI DR2 BAO measurements}",
    eprint = "2503.14743",
    archivePrefix = "arXiv",
    primaryClass = "astro-ph.CO",
    reportNumber = "FERMILAB-PUB-25-0164-PPD",
    month = "3",
    year = "2025"
}

@article{DES:2024jxu,
    author = "Abbott, T. M. C. and others",
    collaboration = "DES",
    title = "{The Dark Energy Survey: Cosmology Results with \ensuremath{\sim}1500 New High-redshift Type Ia Supernovae Using the Full 5 yr Data Set}",
    eprint = "2401.02929",
    archivePrefix = "arXiv",
    primaryClass = "astro-ph.CO",
    reportNumber = "FERMILAB-PUB-23-0821-PPD, DES-2023-805",
    doi = "10.3847/2041-8213/ad6f9f",
    journal = "Astrophys. J. Lett.",
    volume = "973",
    number = "1",
    pages = "L14",
    year = "2024"
}

@misc{DES:2025bxy,
    author = "Abbott, T. M. C. and others",
    collaboration = "DES",
    title = "{Dark Energy Survey: implications for cosmological expansion models from the final DES Baryon Acoustic Oscillation and Supernova data}",
    eprint = "2503.06712",
    archivePrefix = "arXiv",
    primaryClass = "astro-ph.CO",
    reportNumber = "DES-2024-0849, FERMILAB-PUB-25-0127-PPD",
    month = "3",
    year = "2025"
}

@article{Cortes:2024lgw,
    author = "Cort\^es, Marina and Liddle, Andrew R.",
    title = "{Interpreting DESI's evidence for evolving dark energy}",
    eprint = "2404.08056",
    archivePrefix = "arXiv",
    primaryClass = "astro-ph.CO",
    doi = "10.1088/1475-7516/2024/12/007",
    journal = "JCAP",
    volume = "12",
    pages = "007",
    year = "2024"
}

@article{Shlivko:2024llw,
    author = "Shlivko, David and Steinhardt, Paul J.",
    title = "{Assessing observational constraints on dark energy}",
    eprint = "2405.03933",
    archivePrefix = "arXiv",
    primaryClass = "astro-ph.CO",
    doi = "10.1016/j.physletb.2024.138826",
    journal = "Phys. Lett. B",
    volume = "855",
    pages = "138826",
    year = "2024"
}

@article{Luongo:2024fww,
    author = "Luongo, Orlando and Muccino, Marco",
    title = "{Model-independent cosmographic constraints from DESI 2024}",
    eprint = "2404.07070",
    archivePrefix = "arXiv",
    primaryClass = "astro-ph.CO",
    doi = "10.1051/0004-6361/202450512",
    journal = "Astron. Astrophys.",
    volume = "690",
    pages = "A40",
    year = "2024"
}

@article{Yin:2024hba,
    author = "Yin, Wen",
    title = "{Cosmic clues: DESI, dark energy, and the cosmological constant problem}",
    eprint = "2404.06444",
    archivePrefix = "arXiv",
    primaryClass = "hep-ph",
    doi = "10.1007/JHEP05(2024)327",
    journal = "JHEP",
    volume = "05",
    pages = "327",
    year = "2024"
}

@misc{Gialamas:2024lyw,
 archiveprefix = {arXiv},
 author = {Gialamas, Ioannis D. and H\"utsi, Gert and Kannike, Kristjan and Racioppi, Antonio and Raidal, Martti and Vasar, Martin and Veerm\"ae, Hardi},
 eprint = {2406.07533},
 month = {6},
 primaryclass = {astro-ph.CO},
 title = {{Interpreting DESI 2024 BAO: late-time dynamical dark energy or a local effect?}},
 year = {2024}
}

@article{Dinda:2024kjf,
    author = "Dinda, Bikash R.",
    title = "{A new diagnostic for the null test of dynamical dark energy in light of DESI 2024 and other BAO data}",
    eprint = "2405.06618",
    archivePrefix = "arXiv",
    primaryClass = "astro-ph.CO",
    doi = "10.1088/1475-7516/2024/09/062",
    journal = "JCAP",
    volume = "09",
    pages = "062",
    year = "2024"
}

@article{Najafi:2024qzm,
    author = "Najafi, Mahdi and Pan, Supriya and Di Valentino, Eleonora and Firouzjaee, Javad T.",
    title = "{Dynamical dark energy confronted with multiple CMB missions}",
    eprint = "2407.14939",
    archivePrefix = "arXiv",
    primaryClass = "astro-ph.CO",
    doi = "10.1016/j.dark.2024.101539",
    journal = "Phys. Dark Univ.",
    volume = "45",
    pages = "101539",
    year = "2024"
}

@misc{Wang:2024dka,
 archiveprefix = {arXiv},
 author = {Wang, Hao and Piao, Yun-Song},
 eprint = {2404.18579},
 month = {4},
 primaryclass = {astro-ph.CO},
 title = {{Dark energy in light of recent DESI BAO and Hubble tension}},
 year = {2024}
}

@article{Tada:2024znt,
    author = "Tada, Yuichiro and Terada, Takahiro",
    title = "{Quintessential interpretation of the evolving dark energy in light of DESI observations}",
    eprint = "2404.05722",
    archivePrefix = "arXiv",
    primaryClass = "astro-ph.CO",
    doi = "10.1103/PhysRevD.109.L121305",
    journal = "Phys. Rev. D",
    volume = "109",
    number = "12",
    pages = "L121305",
    year = "2024"
}

@article{Carloni:2024zpl,
    author = "Carloni, Youri and Luongo, Orlando and Muccino, Marco",
    title = "{Does dark energy really revive using DESI 2024 data?}",
    eprint = "2404.12068",
    archivePrefix = "arXiv",
    primaryClass = "astro-ph.CO",
    doi = "10.1103/PhysRevD.111.023512",
    journal = "Phys. Rev. D",
    volume = "111",
    number = "2",
    pages = "023512",
    year = "2025"
}

@article{Chan-GyungPark:2024mlx,
    author = "Park, Chan-Gyung and de Cruz P\'erez, Javier and Ratra, Bharat",
    title = "{Using non-DESI data to confirm and strengthen the DESI 2024 spatially flat w0waCDM cosmological parametrization result}",
    eprint = "2405.00502",
    archivePrefix = "arXiv",
    primaryClass = "astro-ph.CO",
    doi = "10.1103/PhysRevD.110.123533",
    journal = "Phys. Rev. D",
    volume = "110",
    number = "12",
    pages = "123533",
    year = "2024"
}

@article{DESI:2024kob,
    author = "Lodha, K. and others",
    collaboration = "DESI",
    title = "{DESI 2024: Constraints on physics-focused aspects of dark energy using DESI DR1 BAO data}",
    eprint = "2405.13588",
    archivePrefix = "arXiv",
    primaryClass = "astro-ph.CO",
    reportNumber = "FERMILAB-PUB-24-0756-PPD",
    doi = "10.1103/PhysRevD.111.023532",
    journal = "Phys. Rev. D",
    volume = "111",
    number = "2",
    pages = "023532",
    year = "2025"
}

@article{Bhattacharya:2024hep,
    author = "Bhattacharya, Sukannya and Borghetto, Giulia and Malhotra, Ameek and Parameswaran, Susha and Tasinato, Gianmassimo and Zavala, Ivonne",
    title = "{Cosmological constraints on curved quintessence}",
    eprint = "2405.17396",
    archivePrefix = "arXiv",
    primaryClass = "astro-ph.CO",
    doi = "10.1088/1475-7516/2024/09/073",
    journal = "JCAP",
    volume = "09",
    pages = "073",
    year = "2024"
}

@article{Ramadan:2024kmn,
    author = "Ramadan, Omar F. and Sakstein, Jeremy and Rubin, David",
    title = "{DESI constraints on exponential quintessence}",
    eprint = "2405.18747",
    archivePrefix = "arXiv",
    primaryClass = "astro-ph.CO",
    doi = "10.1103/PhysRevD.110.L041303",
    journal = "Phys. Rev. D",
    volume = "110",
    number = "4",
    pages = "L041303",
    year = "2024"
}

@misc{Bhattacharya:2024kxp,
    author = "Bhattacharya, Sukannya and Borghetto, Giulia and Malhotra, Ameek and Parameswaran, Susha and Tasinato, Gianmassimo and Zavala, Ivonne",
    title = "{Cosmological tests of quintessence in quantum gravity}",
    eprint = "2410.21243",
    archivePrefix = "arXiv",
    primaryClass = "astro-ph.CO",
    month = "10",
    year = "2024"
}

@article{Orchard:2024bve,
    author = "Orchard, Lili and C\'ardenas, V\'\i{}ctor H.",
    title = "{Probing dark energy evolution post-DESI 2024}",
    eprint = "2407.05579",
    archivePrefix = "arXiv",
    primaryClass = "astro-ph.CO",
    doi = "10.1016/j.dark.2024.101678",
    journal = "Phys. Dark Univ.",
    volume = "46",
    pages = "101678",
    year = "2024"
}

@article{Hernandez-Almada:2024ost,
    author = "Hern\'andez-Almada, A. and Mendoza-Mart\'\i{}nez, M. L. and Garc\'\i{}a-Aspeitia, Miguel A. and Motta, V.",
    title = "{Phenomenological emergent dark energy in the light of DESI Data Release 1}",
    eprint = "2407.09430",
    archivePrefix = "arXiv",
    primaryClass = "astro-ph.CO",
    doi = "10.1016/j.dark.2024.101668",
    journal = "Phys. Dark Univ.",
    volume = "46",
    pages = "101668",
    year = "2024"
}

@misc{Pourojaghi:2024tmw,
    author = "Pourojaghi, S. and Malekjani, M. and Davari, Z.",
    title = "{Cosmological constraints on dark energy parametrizations after DESI 2024: Persistent deviation from standard $\Lambda$CDM cosmology}",
    eprint = "2407.09767",
    archivePrefix = "arXiv",
    primaryClass = "astro-ph.CO",
    month = "7",
    year = "2024"
}

@article{Giare:2024gpk,
 archiveprefix = {arXiv},
 author = {Giar\`e, William and Najafi, Mahdi and Pan, Supriya and Di Valentino, Eleonora and Firouzjaee, Javad T.},
 doi = {10.1088/1475-7516/2024/10/035},
 eprint = {2407.16689},
 journal = {JCAP},
 pages = {035},
 primaryclass = {astro-ph.CO},
 title = {{Robust preference for Dynamical Dark Energy in DESI BAO and SN measurements}},
 volume = {10},
 year = {2024}
}

@article{Reboucas:2024smm,
    author = "Rebou\c{c}as, Jo\~ao and de Souza, Diogo H. F. and Zhong, Kunhao and Miranda, Vivian and Rosenfeld, Rogerio",
    title = "{Investigating late-time dark energy and massive neutrinos in light of DESI Y1 BAO}",
    eprint = "2408.14628",
    archivePrefix = "arXiv",
    primaryClass = "astro-ph.CO",
    doi = "10.1088/1475-7516/2025/02/024",
    journal = "JCAP",
    volume = "02",
    pages = "024",
    year = "2025"
}

@misc{Giare:2024ocw,
 archiveprefix = {arXiv},
 author = {Giar\`e, William},
 eprint = {2409.17074},
 month = {9},
 primaryclass = {astro-ph.CO},
 title = {{Dynamical Dark Energy Beyond Planck? Constraints from multiple CMB probes, DESI BAO and Type-Ia Supernovae}},
 year = {2024}
}

@misc{Chan-GyungPark:2024brx,
    author = "Park, Chan-Gyung and de Cruz Perez, Javier and Ratra, Bharat",
    title = "{Is the $w_0w_a$CDM cosmological parameterization evidence for dark energy dynamics partially caused by the excess smoothing of Planck CMB anisotropy data?}",
    eprint = "2410.13627",
    archivePrefix = "arXiv",
    primaryClass = "astro-ph.CO",
    month = "10",
    year = "2024"
}

@article{Menci:2024hop,
    author = "Menci, Nicola and Sen, Anjan Ananda and Castellano, Marco",
    title = "{The Excess of JWST Bright Galaxies: A Possible Origin in the Ground State of Dynamical Dark Energy in the Light of DESI 2024 Data}",
    eprint = "2410.22940",
    archivePrefix = "arXiv",
    primaryClass = "astro-ph.CO",
    doi = "10.3847/1538-4357/ad8d5b",
    journal = "Astrophys. J.",
    volume = "976",
    number = "2",
    pages = "227",
    year = "2024"
}

@misc{Li:2024qus,
    author = "Li, Tian-Nuo and Li, Yun-He and Du, Guo-Hong and Wu, Peng-Ju and Feng, Lu and Zhang, Jing-Fei and Zhang, Xin",
    title = "{Revisiting holographic dark energy after DESI 2024}",
    eprint = "2411.08639",
    archivePrefix = "arXiv",
    primaryClass = "astro-ph.CO",
    month = "11",
    year = "2024"
}

@misc{Li:2024hrv,
    author = "Li, Jun-Xian and Wang, Shuang",
    title = "{A comprehensive numerical study on four categories of holographic dark energy models}",
    eprint = "2412.09064",
    archivePrefix = "arXiv",
    primaryClass = "astro-ph.CO",
    month = "12",
    year = "2024"
}

@misc{Notari:2024zmi,
    author = "Notari, Alessio and Redi, Michele and Tesi, Andrea",
    title = "{BAO vs. SN evidence for evolving dark energy}",
    eprint = "2411.11685",
    archivePrefix = "arXiv",
    primaryClass = "astro-ph.CO",
    month = "11",
    year = "2024"
}

@article{Gao:2024ily,
    author = "Gao, Qing and Peng, Zhiqian and Gao, Shengqing and Gong, Yungui",
    title = "{On the Evidence of Dynamical Dark Energy}",
    eprint = "2411.16046",
    archivePrefix = "arXiv",
    primaryClass = "astro-ph.CO",
    doi = "10.3390/universe11010010",
    journal = "Universe",
    volume = "11",
    number = "1",
    pages = "10",
    year = "2025"
}

@misc{Fikri:2024klc,
    author = "Fikri, Ramy and ElKhateeb, Esraa and Lashin, El Sayed and El Hanafy, Waleed",
    title = "{A preference for dynamical phantom dark energy using one-parameter model with Planck, DESI DR1 BAO and SN data}",
    eprint = "2411.19362",
    archivePrefix = "arXiv",
    primaryClass = "astro-ph.CO",
    month = "11",
    year = "2024"
}

@misc{Jiang:2024xnu,
 archiveprefix = {arXiv},
 author = {Jiang, Jun-Qian and Pedrotti, Davide and da Costa, Simony Santos and Vagnozzi, Sunny},
 eprint = {2408.02365},
 month = {8},
 primaryclass = {astro-ph.CO},
 title = {{Non-parametric late-time expansion history reconstruction and implications for the Hubble tension in light of DESI}},
 year = {2024}
}

@misc{Zheng:2024qzi,
    author = "Zheng, Jie and Qiang, Da-Chun and You, Zhi-Qiang",
    title = "{Cosmological constraints on dark energy models using DESI BAO 2024}",
    eprint = "2412.04830",
    archivePrefix = "arXiv",
    primaryClass = "astro-ph.CO",
    month = "12",
    year = "2024"
}

@article{Gomez-Valent:2024ejh,
    author = "G\'omez-Valent, Adria and Sol\`a Peracaula, Joan",
    title = "{Composite dark energy and the cosmological tensions}",
    eprint = "2412.15124",
    archivePrefix = "arXiv",
    primaryClass = "astro-ph.CO",
    doi = "10.1016/j.physletb.2025.139391",
    journal = "Phys. Lett. B",
    volume = "864",
    pages = "139391",
    year = "2025"
}

@article{RoyChoudhury:2024wri,
    author = "Roy Choudhury, Shouvik and Okumura, Teppei",
    title = "{Updated Cosmological Constraints in Extended Parameter Space with Planck PR4, DESI Baryon Acoustic Oscillations, and Supernovae: Dynamical Dark Energy, Neutrino Masses, Lensing Anomaly, and the Hubble Tension}",
    eprint = "2409.13022",
    archivePrefix = "arXiv",
    primaryClass = "astro-ph.CO",
    doi = "10.3847/2041-8213/ad8c26",
    journal = "Astrophys. J. Lett.",
    volume = "976",
    number = "1",
    pages = "L11",
    year = "2024"
}

@misc{RoyChoudhury:2025dhe,
    author = "Roy Choudhury, Shouvik",
    title = "{Cosmology in Extended Parameter Space with DESI DR2 BAO: A 2$\sigma$+ Detection of Non-zero Neutrino Masses with an Update on Dynamical Dark Energy and Lensing Anomaly}",
    eprint = "2504.15340",
    archivePrefix = "arXiv",
    primaryClass = "astro-ph.CO",
    month = "4",
    year = "2025"
}

@misc{Lewis:2024cqj,
    author = "Lewis, Antony and Chamberlain, Ewan",
    title = "{Understanding acoustic scale observations: the one-sided fight against $\Lambda$}",
    eprint = "2412.13894",
    archivePrefix = "arXiv",
    primaryClass = "astro-ph.CO",
    month = "12",
    year = "2024"
}

@misc{Wolf:2025jlc,
    author = "Wolf, William J. and Garc\'\i{}a-Garc\'\i{}a, Carlos and Ferreira, Pedro G.",
    title = "{Robustness of Dark Energy Phenomenology Across Different Parameterizations}",
    eprint = "2502.04929",
    archivePrefix = "arXiv",
    primaryClass = "astro-ph.CO",
    month = "2",
    year = "2025"
}

@misc{Shajib:2025tpd,
    author = "Shajib, Anowar J. and Frieman, Joshua A.",
    title = "{Evolving dark energy models: Current and forecast constraints}",
    eprint = "2502.06929",
    archivePrefix = "arXiv",
    primaryClass = "astro-ph.CO",
    month = "2",
    year = "2025"
}

@article{Giare:2025pzu,
    author = "Giar\`e, William and Mahassen, Tariq and Di Valentino, Eleonora and Pan, Supriya",
    title = "{An overview of what current data can (and cannot yet) say about evolving dark energy}",
    eprint = "2502.10264",
    archivePrefix = "arXiv",
    primaryClass = "astro-ph.CO",
    doi = "10.1016/j.dark.2025.101906",
    journal = "Phys. Dark Univ.",
    volume = "48",
    pages = "101906",
    year = "2025"
}

@misc{Chaussidon:2025npr,
    author = "Chaussidon, E. and others",
    title = "{Early time solution as an alternative to the late time evolving dark energy with DESI DR2 BAO}",
    eprint = "2503.24343",
    archivePrefix = "arXiv",
    primaryClass = "astro-ph.CO",
    month = "3",
    year = "2025"
}

@misc{Kessler:2025kju,
    author = "Kessler, Daniel A. and Escamilla, Luis A. and Pan, Supriya and Di Valentino, Eleonora",
    title = "{One-parameter dynamical dark energy: Hints for oscillations}",
    eprint = "2504.00776",
    archivePrefix = "arXiv",
    primaryClass = "astro-ph.CO",
    month = "4",
    year = "2025"
}

@misc{Alvarez:2025kma,
    author = "Alvarez, Marco Antonio Cardoso and Graef, Leila and Brandenberger, Robert",
    title = "{Back-Reaction of Super-Hubble Fluctuations, Late Time Tracking and Recent Observational Results}",
    eprint = "2502.17395",
    archivePrefix = "arXiv",
    primaryClass = "astro-ph.CO",
    month = "2",
    year = "2025"
}

@article{Chevallier:2000qy,
    author = "Chevallier, Michel and Polarski, David",
    title = "{Accelerating universes with scaling dark matter}",
    eprint = "gr-qc/0009008",
    archivePrefix = "arXiv",
    doi = "10.1142/S0218271801000822",
    journal = "Int. J. Mod. Phys. D",
    volume = "10",
    pages = "213--224",
    year = "2001"
}

@article{Linder:2002et,
    author = "Linder, Eric V.",
    title = "{Exploring the expansion history of the universe}",
    eprint = "astro-ph/0208512",
    archivePrefix = "arXiv",
    doi = "10.1103/PhysRevLett.90.091301",
    journal = "Phys. Rev. Lett.",
    volume = "90",
    pages = "091301",
    year = "2003"
}

@article{Ozulker:2022slu,
    author = "Ozulker, Emre",
    title = "{Is the dark energy equation of state parameter singular?}",
    eprint = "2203.04167",
    archivePrefix = "arXiv",
    primaryClass = "astro-ph.CO",
    doi = "10.1103/PhysRevD.106.063509",
    journal = "Phys. Rev. D",
    volume = "106",
    number = "6",
    pages = "063509",
    year = "2022"
}

@article{Sussman:2024mzs,
    author = "Sussman, Roberto A. and N{\'a}jera, Sebasti{\'a}n",
    title = "{Schouten{\textendash}Codazzi gravity}",
    eprint = "2411.06257",
    archivePrefix = "arXiv",
    primaryClass = "gr-qc",
    doi = "10.1088/1361-6382/ad922e",
    journal = "Class. Quant. Grav.",
    volume = "41",
    number = "24",
    pages = "245019",
    year = "2024"
}

@article{Harada:2021cotton,
  author  = {Harada, Junya},
  title   = {Gravity with a trace of {C}otton},
  journal = {Phys.\ Rev.\ D},
  volume  = {103},
  pages   = {124034},
  year    = {2021},
  doi     = {10.1103/PhysRevD.103.124034},
  eprint  = {2104.00571},
  archivePrefix = {arXiv},
  primaryClass  = {gr-qc}
}

@article{Mantica:2022flg,
    author = "Mantica, Carlo Alberto and Molinari, Luca Guido",
    title = "{Codazzi tensors and their space-times and Cotton gravity}",
    eprint = "2210.06173",
    archivePrefix = "arXiv",
    primaryClass = "gr-qc",
    doi = "10.1007/s10714-023-03106-7",
    journal = "Gen. Rel. Grav.",
    volume = "55",
    number = "4",
    pages = "62",
    year = "2023"
}

@article{Rubin:2023jdq,
    author = "Rubin, David and others",
    title = "{Union Through UNITY: Cosmology with 2,000 SNe Using a Unified Bayesian Framework}",
    eprint = "2311.12098",
    archivePrefix = "arXiv",
    primaryClass = "astro-ph.CO",
    doi = "10.3847/1538-4357/adc0a5",
    journal = "Astrophys. J.",
    volume = "986",
    number = "2",
    pages = "231",
    year = "2025"
}

@article{Buchner:2021ultranest,
  author  = {Buchner, Johannes},
  title   = {{UltraNest} --- a robust, general purpose {B}ayesian inference engine},
  journal = {J.\ Open Source Softw.},
  volume  = {6},
  number  = {60},
  pages   = {3001},
  year    = {2021},
  doi     = {10.21105/joss.03001}
}

\end{document}